\documentclass{article}
\usepackage{graphicx}
\usepackage[bib=library]{mathesis}
\title{Pricing American Options by Exercise Rate Optimization}
\author{
	{\large Christian Bayer\textsuperscript{a}, Raúl Tempone\textsuperscript{b,c}, Sören Wolfers\textsuperscript{c}\textsuperscript{,}\footnote{Corresponding author. Email address: \texttt{soeren.wolfers@kaust.edu.sa}}}\\[0.5em]
	{\small \textsuperscript{a} Weierstrass Institute for Applied Analysis and Stochastics (WIAS), Berlin, Germany}\\
	{\small \textsuperscript{b} RWTH Aachen University, Germany}\\
	{\small \textsuperscript{c} King Abdullah University of Science and Technology (KAUST), Thuwal, Saudi Arabia}\\
}
\date{\today}

\renewcommand{\d}{\mathop{}\!\mathrm{d}}
\newcommand{\payoff}{g}
\newcommand{\T}{\mathcal{T}}
\usepackage{siunitx}
\usepackage{todonotes}
\newcommand{\variance}{v}

\usepackage{hhline}
\begin{document}
\maketitle

\textbf{Abstract}
We present a novel method for the numerical pricing of American options based on Monte Carlo simulation and the optimization of exercise strategies.
Previous solutions to this problem either explicitly or implicitly determine so-called optimal \emph{exercise regions}, which consist of points in time and space at which a given option is exercised. 
In contrast, our method determines the \emph{exercise rates} of randomized exercise strategies. We show that the supremum of the corresponding stochastic optimization problem provides the correct option price. 
By integrating analytically over the random exercise decision, we obtain an objective function that is differentiable with respect to perturbations of the exercise rate even for finitely many sample paths. 
The global optimum of this function can be approached gradually when starting from a constant exercise rate. 
 Numerical experiments on vanilla put options in the multivariate Black--Scholes model and a preliminary theoretical analysis underline the efficiency of our method, both with respect to the number of time-discretization steps and the required number of degrees of freedom in the parametrization of the exercise rates. 
Finally, we demonstrate the flexibility of our method through numerical experiments on max call options in the classical Black--Scholes model, and vanilla put options in both the Heston model and the non-Markovian rough Bergomi model.
\\ 

{\small
	
}
 
\section{Introduction}
American options on $d\geq 1$ underlying assets
$S_{t}=(S_{1,t},\dots,S_{d,t})$  may be exercised by their holder at any time $t$ before a given expiration time $T\in\R_+:=[0,\infty)$, upon which
the holder receives the payoff $\payoff (t,S_t)$ for some previously agreed
function $\payoff\colon  [0,T] \times \R_{+}^{d} \to \R_{+}$.
 
If the underlying market is Markovian and has a security with interest rate
$r>0$, then the  arbitrage-free value of an American option under a risk-neutral 
measure $\mathbb{Q}$ is determined solely by the
current asset values. %Moreover, if we assume for simplicity that the
%underlying assets do not pay dividends, then t
The value function $V\colon \R_{+}^{d}\to\R_{+}$ satisfies
\begin{equation}
\label{fundstop}
V(s_0)=\sup_{\tau \in \mathcal{S}}\E_{\mathbb{Q}} [Y_{\tau\wedge
  T}|{S}_{0}=s_0],\quad s_0\in \R_{+}^{d},
\end{equation}
where $Y_{t}:=\exp(-rt)\payoff (t, S_{t})$, $t\geq 0$ is the {discounted payoff process} and $\mathcal{S}$ denotes the set of all stopping times with respect to the filtration generated by $(S_t)_{0\leq t\leq T}$  \cite[Theorem 5.3]{karatzas1998methods}. In the remainder of this work, all expectations are taken with respect to the same risk-neutral measure $\mathbb{Q}$ and denoted by $\E$. 

Most state-of-the-art methods for American option pricing -- including all variants of the Longstaff--Schwartz~\cite{longstaff2001valuing}, PDE~\cite{achdou2005computational}, binomial tree~\cite{cox1979option}, and stochastic mesh~\cite{broadie1997pricing} methods -- exploit the dynamic
programming principle to determine the value function using a backwards-iteration scheme. Further approaches are based on dual problems~\cite{rogers2002montecarlo,andersen2004primal}, policy iteration~\cite{belomestny2018advanced}, or (quasi-)analytic solutions~\cite{barone1987efficient,kuske1998optimal}. 
The computational cost of many methods grows exponentially with respect to the number of dimensions, thus making them prohibitively expensive for options on many underlying assets. This phenomenon has been coined the curse of dimensionality \cite{reisinger2007efficient,bellman2015adaptive}.

In this work, we propose a method that is based on the following variation of \Cref{fundstop}, which states that the optimization may be restricted to \emph{hitting times} instead of general stopping times:
\begin{equation}
\label{fund}
V(s_0)=\sup_{E\in \mathcal{B}([0,T]\times \R_{+}^{d})}\E [Y_{\tau_{E}\wedge T}|{S}_{0}=s_0],\quad s_0\in \R_{+}^{d}.
\end{equation}
Here, the supremum is taken over Borel-measurable subsets of $E\subset[0,T]\times \R_{+}^{d}$, whose {hitting times} are given by  $\tau_{E}:=\inf\{t\geq 0 :
(t,S_t)\in E\}$.  
To be precise, both \Cref{fundstop} and \Cref{fund} require some technical conditions on the processes $(Y_{t})_{0\leq t\leq T}$ and $(S_t)_{0\leq t \leq T}$ \cite[Corollary 2, Section 3.3.1]{shiryaev2007optimal}. Throughout this work, we assume that such conditions hold and restrict our attention to the solution of \Cref{fund}.

To the best of our knowledge, optimization of the exercise region in \Cref{fund} was first proposed in~\cite{grant1997path} and developed in \cite{andersen1999simple,garcia2003convergence,ibanez2004monte,belomestny2011on,gemmrich2012master}, but it has not yet found its way into the canon of numerical algorithms for American option pricing. In~\cite{grant1997path}, separate exercise regions were determined for each exercise date of an American Asian option in a backwards iteration. The optimization at each step was performed in a brute force fashion, which explains why only two parameters were allowed in the parametrization of the exercise regions. In \cite{garcia2003convergence,gemmrich2012master}, ad hoc parametrizations that exploit known behavior of the optimal exercise regions were used to optimize exercise regions as subsets of time-space without applying a backwards iteration.

In general, optimization of the exercise region faces two challenges. First, as mentioned in \cite{gemmrich2012master}, it is not obvious how to parametrize the possible exercise regions in a multi-dimensional setting, or even in a one-dimensional setting that goes beyond vanilla options in the Black--Scholes model. Second, once a parametrization has been found,
 it is not obvious how to find the global optimum \cite{garcia2003convergence,gemmrich2012master}. Indeed, when the expectation in \Cref{fund} is replaced by an empirical average for the purpose of numerical approximations of the expected payoff, the quantity to be maximized depends highly irregularly on the exercise region $E$ (see \Cref{fig:psi1theirs} below).
Furthermore, even if a large number of sample paths is used to reduce the small scale oscillatory behavior, the resulting surface may still be non-concave and exhibit isolated local optima, as reported in \cite{garcia2003convergence}. 

To address these challenges, we introduce, in \Cref{sec:ero}, a relaxation of the
optimization problem in \Cref{fund} wherein the exercise regions $E\subset
[0,T] \times \R_{+}^{d}$ are replaced by \emph{exercise rates} $f\colon[0,T] \times \R_{+}^{d}\to\R_{+}$, which define \emph{randomized exercise strategies} where options are exercised with an infinitesimal probability depending on the current time and asset values.\footnote{We were informed after the initial submission of this manuscript that randomized stopping was previously studied from a theoretical perspective \cite{gyongy2008randomized,krylov2008controlled}. These references do not contain discussions of numerical solution of the resulting stochastic optimization problem, however.}
The space of exercise rates can easily be parametrized even in high dimensions using a finite-dimensional spaces of polynomials on $[0,T]\times \R^{d}$.
The resulting optimization problem exhibits the same maximum as the original optimization problem over deterministic strategies but has the advantage of a differentiable objective function and a lower risk of getting stuck in local minima because of a richer search space. 
Indeed, by integrating analytically with respect to the exponential distribution that underlies the random exercise decision, we obtain an objective function that is smooth even when finitely many sample paths are used in the computations. We may then use gradient-based optimization routines to determine an optimal coefficient vector. Furthermore, we may start this optimization from an exercise rate that has a constant non-zero value across time and space and let the optimization routine gradually refine this neutral strategy towards an optimal one with marked variations in the exercise rate. This facilitates the search for a global optimum without requiring an informed initial guess that is already close to the optimum. Details of the numerical implementation are discussed in \Cref{sec:algo}. 
There, we also briefly discuss how the accuracy of our method depends on the various discretization parameters. In particular, we provide heuristic bounds on the number of degrees of freedom in the exercise rate that are required for satisfactory randomized exercise strategies. These bounds are given in terms of the smoothness of the optimal exercise boundary as a manifold, not as a function of time.

Finally, \Cref{sec:practice} presents numerical experiments for various market models and options. 
In \Cref{sec:practice0,sec:practice1}, we consider vanilla put options in the classical Black--Scholes model. 
In the case of a single underlying, the exercise boundary of an American put option, whose payoff function is given by $\payoff (t,s):=\payoff (s):=(K-s)^{+}$ for some {strike} $K>0$, can be written as a function of time with asymptotic behavior $s(t)\approx K-C_1\sqrt{(T-t)\log (T-t)}$ for some $C_1>0$ as $t\to T$. Despite the square-root singularity near the expiration time, the experiments presented in \Cref{sec:practice0} show that low-degree polynomials suffice to capture the optimal exercise boundary well. In fact, we obtain a relative error of less than $0.1\%$ with quadratic polynomials. This can be explained by the fact that the graph of the similar function $\tilde{s}(t)=K-C_1\sqrt{(T-t)}$ is smooth as a one-dimensional manifold in $\R^2$ and, indeed, coincides with the zero level set (intersected with $x<K$) of the quadratic polynomial $f(t,s):=(K-s)^2-C_1^2(T-t)$, whose scalar multiples therefore constitute close-to-optimal exercise rates.

 Although we  solve non-concave maximization problems, we are able to find global optima starting from a constant exercise rate. %Since we do not perform backwards iteration in time, our method is insensitive to the number of exercise dates. This is in contrast to the exponential increase of the error with respect to the number of exercise dates in conventional Longstaff--Schwartz algorithms \cite{zanger2018convergence}. 
Furthermore, in \Cref{sec:practice1} we show that our algorithm outperforms the Longstaff--Schwartz algorithm with respect to the required polynomial degree for the pricing of basket put options, which is crucial when the number of underlying asset is large. 

In \Cref{sec:practice1b}, we consider call options on the maximum of a number of underlying assets, $g(s)=\max_{i=1}^{d}(s_i-K)^{+}$. Numerical algorithms for the pricing of such max call options were previously discussed in   \cite{andersen2004primal,ludkovski2018kriging}. Max call options pose a challenge to the direct determination of exercise regions because the optimal exercise regions are disconnected \cite{broadie1997valuation}. Still, our results show that polynomials of low degree suffice to obtain highly accurate estimates despite the nontrivial topology of the optimal exercise region.

In \Cref{sec:practice2}, we consider the {Heston} model, in which the underlying asset and its stochastic volatility form a joint Markov process. Since our method involves the market model for the generation of random sample paths only, its application in this scenario is straightforward.
Finally, we consider the non-Markovian {rough Bergomi} model \cite{bayer2016pricing} in \Cref{sec:practice3}. To recover Markovianity, we must extend our process by its past values. In practice, using a large but finite number of past values leads to very high-dimensional approximation problems. However, our experiments indicate that exercise strategies depending only on the spot values of the underlying asset and its volatility achieve near-optimal performance.

\section{Exercise rate optimization}
\label{sec:ero}
We let  $\T:=[0,T]$ and assume throughout that $(S_{t})_{t\in \T}$ is conditioned on $S_{0}=s_0$.
\begin{definition}
For any $f\colon\T\times \R_{+}^{d}\to\R_{+}$, the \emph{randomized exercise strategy with exercise rate} $f$ is given by early exercise at the time 
\begin{equation}
\label{randomstop}
\tau_{f}:=\inf\{t\geq 0:\int_{0}^{t}\lambda_u\d u \geq X\},
\end{equation} where 
$
\lambda_{t}:=f(t,S_t)
$, $t\in\T$, and $X$ is a standard exponential random variable that is independent of $(S_t)_{t\in \T}$. 
\end{definition}
 The exercise time $\tau_f$ equals the first jump time of a Poisson process with rate $(\lambda_{t})_{t\in \T}$. In other words, the exercise rate $f$ determines the time- and space-dependent infinitesimal probability with which the American option is exercised in a infinitesimal time interval $\d t$.
 
 With \Cref{fund} in mind, we are interested in the expected payoff under a randomized exercise strategy with early exercise time $\tau_{f}$, which we denote by
 \begin{equation}
 \label{psi}
\psi(f):=\E [Y_{\tau_{f}\wedge T}].
 \end{equation}
  Since $\int_{0}^{t}\lambda_u\d u$ is a deterministic function of the asset path until $t$, and $X$  is independent of $(S_u)_{u\in \T}$, we have
  $$
  \mathbb{P}(\tau_f \geq t\mid (S_{u})_{u\in \T})	=\mathbb{P}(X>\int_{0}^{t}\lambda_u\d u\mid (S_{u})_{u\in \T})=\exp\left(-\int_{0}^{t}\lambda_u \d u\right)=:U_{t}
  $$ and 
  $$
  \mathbb{P}(\tau_f \in \d t\mid (S_{u})_{u\in\T})= -\d U_t
 = \lambda_t U_t \d t .
 $$
 Hence, we obtain 
 \begin{equation*} 
 \phi(f,(S_{u})_{u\in\T}):=\E [Y_{\tau_{f}\wedge T}\mid (S_{u})_{u\in\T}] = \int_{0}^T Y_t \lambda_t U_t \d t +
 Y_T U_T.
 \end{equation*}
By the law of total expectation, which we may apply because all the random variables involved are nonnegative, we deduce the formula
  	\begin{equation}
  	\label{fund1}
  	\psi(f)= \E[\phi(f,(S_{u})_{u\in\T})] = \E\left[\int_{0}^T Y_t \lambda_t U_t \d t +
  	Y_T U_T\right].
  	\end{equation}
	It is advisable to replace $\lambda U_t \d t$ by $-\d U_t$ in numerical implementations of this formula to avoid cancellations.
The following proposition shows that, in theory, exercise rate optimization yields the correct option value. It is a special case of Theorem 2.2 in \cite{gyongy2008randomized}.
\begin{proposition}
	\label{equality}
	We have
	\begin{equation}
	\label{fund3}
	V(s_0)=\sup_{f\colon [0,T]\times \R_{+}^{d}\to \R_{+}}\psi(f).
	\end{equation}
\end{proposition}
\begin{proof}
	For any $E\in \mathcal{B}(\T\times \R_{+}^{d})$, we may formally insert the indicator function
	\begin{equation*}
	f_{E}(t,s):=\begin{cases}
	+\infty, & (t,s)\in E\\
	0, &(t,s)\not\in E
	\end{cases}
	\end{equation*} 
	into \Cref{randomstop} to obtain $\tau_{f_E}=\tau_E$. After replacing $+\infty$ 
	with large numbers that diverge to $+\infty$ and applying Fatou's lemma, we may take the supremum over $E$ to conclude from \Cref{fund}
        that $\sup_{f\colon [0,T] \times \R_{+}^{d}\to \R_{+}}\psi(f)\geq V(s_0)$.
	
	Conversely, the law of total expectation shows, for any $f\colon [0,T]
        \times \R_{+}^{d}\to\R_{+}$, that
	\begin{align*}
	\psi(f)=\E[Y_{\tau_{f}\wedge T}]=\E\Big[\E\left[Y_{\tau_f\wedge T}\mid X\right]\Big].
	\end{align*}
	Because $\tau_{f}$ conditioned on $X$ is a stopping time and $(S_{t})_{t\in\T}$ is independent of $X$, \Cref{fundstop} implies that $\E\left[Y_{\tau_{f}\wedge T}\mid X\right]\leq V(s_0)$ almost surely; hence, $\psi(f)\leq V(s_0)$. 
\end{proof}
\subsection{Numerical algorithm}
\label{sec:algo} 
To determine optimal exercise rates numerically, we 
\begin{enumerate}[(i)]
\item replace the time-continuous model of the stochastic process $(S_{t})_{t\in\T}$ with a discretization with $N<\infty$ time steps, such as the the Euler--Maruyama scheme;
  
\item  replace the expectation in \Cref{fund1} with an average over $M<\infty$ fixed sample paths $(S^{(m)}_{n})_{1\leq n\leq N,1\leq m \leq M}$;
  
\item \label{pparam} introduce a $B$-dimensional, $B<\infty$ parametrization $\R^{B}\ni\bm{c}\mapsto f_{\bm{c}}$ of the space of exercise rates;
  
\item maximize the surrogate function
  \begin{align*}
    \overline{\psi}\colon &\R^{B}\to \R\\
                          &\bm{c}\mapsto \frac{1}{M}\sum_{m=1}^{M} \phi(f_{\bm{c}},(S^{(m)}_{t})_{1\leq n\leq N}).\footnotemark
  \end{align*} \footnotetext{To evaluate $\phi$, we use piecewise constant interpolation between the $N$ nodes of the time-discretization scheme.}
\end{enumerate} 

\paragraph{Parametrization} To address step (iii), we work with the logarithmic asset values $x_i:=\log(s_i)$, ${1\leq i\leq d}$ and let
\begin{equation*}
F_{\mathcal{P}}:=\left\{f_{p}(t,x):=1_{\payoff (t,s)>0}\exp(p(t,x))\;\big|\;p\in \mathcal{P}\right\}
\end{equation*} 
for any finite-dimensional linear space $\mathcal{P}$ of functions on $\T\times\R^{d}$. After choosing a basis of $\mathcal{P}$, we obtain the desired parametrization $\bm{c}\mapsto f_{\bm{c}}$. Throughout the remainder of this manuscript, we work with spaces $\mathcal{P}_{k}$ of polynomials of degree less than or equal to $k\geq 0$ in $d+1$ variables, and we use an orthonormal basis with respect to the inner product $\|f\|^2:=\frac{1}{NM}\sum_{n=1}^{N}\sum_{m=1}^{M} f(t_n,x_{n,m})$ induced by the time-space samples $(t_{n},x_{n,m}:=\log(S_{n}^{m}))_{1\leq n\leq N, 1\leq m\leq M}$.

\paragraph{Optimization} Concerning step (iv), it is not clear that globally optimal coefficients, which may even lie at infinity, can be found numerically because $\overline{\psi}$ is not concave. However, in our numerical experiments, we found that the Quasi-Newton L-BFGS-B algorithm \cite{byrd1995limited}, as implemented in Python's SciPy library\footnote{\url{https://docs.scipy.org/doc/scipy/reference/optimize.minimize-lbfgsb.html}}, performs well and does not get stuck in local maxima when started from a constant exercise rate. 

\begin{figure}[h!]
	\begin{subfigure}{0.49\textwidth}
		\scalebox{0.6}{% This file was created by matplotlib2tikz v0.6.17.
\begin{tikzpicture}

\definecolor{color0}{rgb}{0.12156862745098,0.466666666666667,0.705882352941177}

\begin{axis}[
width=14cm,
height=8cm,
xlabel={\large $c_{(0,0)}$},
xmin=-10, xmax=10,
ymin=0, ymax=10,
yticklabel style = {font=\large,xshift=0.5ex},
xticklabel style = {font=\large,yshift=0.5ex},
tick align=outside,
max space between ticks=1000pt,
try min ticks=5,
tick pos=left,
x grid style={white!69.01960784313725!black},
y grid style={white!69.01960784313725!black},
]
\addplot [semithick, color0, forget plot]
table[y expr=\thisrowno{1}*100] {%
-9.9 0.0583127135211175
-9.80050251256282 0.0583128288412495
-9.70100502512563 0.0583129788531234
-9.60150753768844 0.058313174203156
-9.50201005025126 0.0583134279891144
-9.40251256281407 0.0583137560641118
-9.30301507537689 0.058314177315148
-9.2035175879397 0.0583147139005562
-9.10402010050251 0.0583153914307414
-9.00452261306533 0.0583162390775384
-8.90502512562814 0.0583172895993737
-8.80552763819096 0.0583185792721036
-8.70603015075377 0.058320147718778
-8.60653266331658 0.0583220376354557
-8.5070351758794 0.0583242944143433
-8.40753768844221 0.0583269656697039
-8.30804020100503 0.0583301006759494
-8.20854271356784 0.0583337497308742
-8.10904522613065 0.0583379634599311
-8.00954773869347 0.0583427920796552
-7.91005025125628 0.0583482846397346
-7.8105527638191 0.0583544882637669
-7.71105527638191 0.0583614474084652
-7.61155778894472 0.0583692031600311
-7.51206030150754 0.0583777925847117
-7.41256281407035 0.0583872481483171
-7.31306532663317 0.0583975972168351
-7.21356783919598 0.0584088616473871
-7.11407035175879 0.0584210574757681
-7.01457286432161 0.0584341947038222
-6.91507537688442 0.0584482771870558
-6.81557788944724 0.0584633026202613
-6.71608040201005 0.0584792626165996
-6.61658291457286 0.058496142873614
-6.51708542713568 0.0585139234180551
-6.41758793969849 0.0585325789201901
-6.31809045226131 0.0585520790674562
-6.21859296482412 0.0585723889868568
-6.11909547738694 0.058593469705393
-6.01959798994975 0.0586152786380566
-5.92010050251256 0.0586377700936977
-5.82060301507538 0.0586608957912994
-5.72110552763819 0.0586846053857722
-5.62160804020101 0.0587088470209122
-5.52211055276382 0.0587335679732228
-5.42261306532663 0.0587587155483774
-5.32311557788945 0.0587842385681209
-5.22361809045226 0.0588100900459321
-5.12412060301508 0.0588362319514845
-5.02462311557789 0.0588626431913735
-4.9251256281407 0.0588893319109435
-4.82562814070352 0.0589163527745124
-4.72613065326633 0.0589438289305541
-4.62663316582915 0.0589719770162266
-4.52713567839196 0.0590011321018163
-4.42763819095477 0.0590317683445035
-4.32814070351759 0.0590645107184509
-4.2286432160804 0.059100133751467
-4.12914572864322 0.0591395447017408
-4.02964824120603 0.0591837507653299
-3.93015075376884 0.0592338122615487
-3.83065326633166 0.0592907858225856
-3.73115577889447 0.059355663102048
-3.63165829145729 0.0594293115583888
-3.5321608040201 0.0595124255937831
-3.43266331658291 0.0596055013446817
-3.33316582914573 0.0597088598195573
-3.23366834170854 0.0598227600407985
-3.13417085427136 0.0599476553871755
-3.03467336683417 0.0600846313258456
-2.93517587939698 0.0602360040495443
-2.8356783919598 0.0604059658942256
-2.73618090452261 0.0606010771852095
-2.63668341708543 0.0608303804269034
-2.53718592964824 0.0611049817314936
-2.43768844221105 0.0614370877140844
-2.33819095477387 0.0618386460165146
-2.23869346733668 0.0623198504173625
-2.1391959798995 0.0628877998476302
-2.03969849246231 0.0635455460960991
-1.94020100502513 0.0642916586407302
-1.84070351758794 0.0651203177661559
-1.74120603015075 0.0660218521089588
-1.64170854271357 0.0669835809948114
-1.54221105527638 0.0679908065506621
-1.4427135678392 0.0690278163728731
-1.34321608040201 0.0700787913503664
-1.24371859296482 0.0711285528483898
-1.14422110552764 0.0721631200604589
-1.04472361809045 0.0731700771061249
-0.945226130653266 0.0741387688660019
-0.845728643216081 0.0750603552704343
-0.746231155778894 0.0759277575828894
-0.646733668341708 0.0767355292502825
-0.547236180904523 0.0774796800638837
-0.447738693467336 0.0781574772168743
-0.348241206030151 0.0787672414324399
-0.248743718592964 0.0793081513343397
-0.149246231155779 0.0797800649842277
-0.0497487437185935 0.0801833641346258
0.0497487437185935 0.0805188242176004
0.149246231155779 0.080787511300081
0.248743718592966 0.0809907060427935
0.348241206030151 0.0811298539452867
0.447738693467336 0.0812065406922719
0.547236180904523 0.0812224910939035
0.646733668341708 0.0811795898035708
0.746231155778895 0.0810799215818372
0.845728643216081 0.080925828248186
0.945226130653266 0.0807199785357234
1.04472361809045 0.0804654457806896
1.14422110552764 0.08016578673142
1.24371859296483 0.0798251128223737
1.34321608040201 0.0794481432165718
1.4427135678392 0.0790402271190346
1.54221105527638 0.0786073218358569
1.64170854271357 0.0781559135157196
1.74120603015075 0.0776928703042653
1.84070351758794 0.0772252235724016
1.94020100502513 0.076759882430211
2.03969849246231 0.0763032996580301
2.1391959798995 0.0758611220266076
2.23869346733668 0.0754378717490961
2.33819095477387 0.0750367140444738
2.43768844221105 0.074659363368073
2.53718592964824 0.0743061637531383
2.63668341708543 0.0739763464709457
2.73618090452261 0.0736684263091205
2.8356783919598 0.0733806580483399
2.93517587939698 0.0731114525872142
3.03467336683417 0.0728596600511373
3.13417085427136 0.0726246668165985
3.23366834170854 0.0724063115579691
3.33316582914573 0.0722046784164406
3.43266331658291 0.0720198510605402
3.5321608040201 0.0718517023342245
3.63165829145729 0.0716997615970473
3.73115577889447 0.071563166573931
3.83065326633166 0.0714406843266384
3.93015075376884 0.0713307795622253
4.02964824120603 0.0712317111253603
4.12914572864322 0.0711416416601862
4.2286432160804 0.0710587479279035
4.32814070351759 0.0709813206692237
4.42763819095477 0.0709078446574353
4.52713567839196 0.0708370524525517
4.62663316582915 0.0707679492229989
4.72613065326633 0.0706998101918691
4.82562814070352 0.0706321559334487
4.9251256281407 0.0705647131290096
5.02462311557789 0.0704973690613764
5.12412060301508 0.070430127130259
5.22361809045226 0.0703630684936114
5.32311557788945 0.0702963223288181
5.42261306532663 0.070230044889697
5.52211055276382 0.0701644059872217
5.62160804020101 0.0700995808645605
5.72110552763819 0.0700357454921177
5.82060301507538 0.0699730737550888
5.92010050251256 0.0699117355447216
6.01959798994975 0.0698518952103963
6.11909547738694 0.0697937101226069
6.21859296482412 0.0697373292567789
6.31809045226131 0.0696828917819493
6.41758793969849 0.0696305256681784
6.51708542713568 0.0695803463375803
6.61658291457286 0.0695324553879895
6.71608040201005 0.0694869394199935
6.81557788944724 0.0694438689986299
6.91507537688442 0.0694032977806973
7.01457286432161 0.0693652618373113
7.1140703517588 0.0693297791989887
7.21356783919598 0.0692968496470772
7.31306532663317 0.0692664547707767
7.41256281407035 0.0692385583033198
7.51206030150754 0.0692131067441974
7.61155778894472 0.0691900302667642
7.71105527638191 0.0691692439023499
7.81055276381909 0.0691506489834061
7.91005025125628 0.0691341348195642
8.00954773869347 0.0691195805721365
8.10904522613066 0.0691068572849739
8.20854271356784 0.069095830023106
8.30804020100503 0.0690863600656386
8.40753768844221 0.0690783070963283
8.5070351758794 0.0690715313343756
8.60653266331658 0.0690658955494695
8.70603015075377 0.0690612669090334
8.80552763819095 0.0690575186118904
8.90502512562814 0.0690545312709461
9.00452261306533 0.0690521940176128
9.10402010050251 0.0690504053120408
9.2035175879397 0.0690490734551817
9.30301507537689 0.0690481168105971
9.40251256281407 0.0690474637550512
9.50201005025126 0.069047052386642
9.60150753768844 0.0690468300269727
9.70100502512563 0.0690467525592326
9.80050251256281 0.0690467836468178
9.9 0.069046893877246
};
\end{axis}

\end{tikzpicture}}
		\caption{$\overline{\psi}\colon\R\to\R$ with $\mathcal{P}$ the space of constant functions}
		\label{fig:psi1ours}
	\end{subfigure}
	\begin{subfigure}{0.49\textwidth}
		\scalebox{0.6}{\input{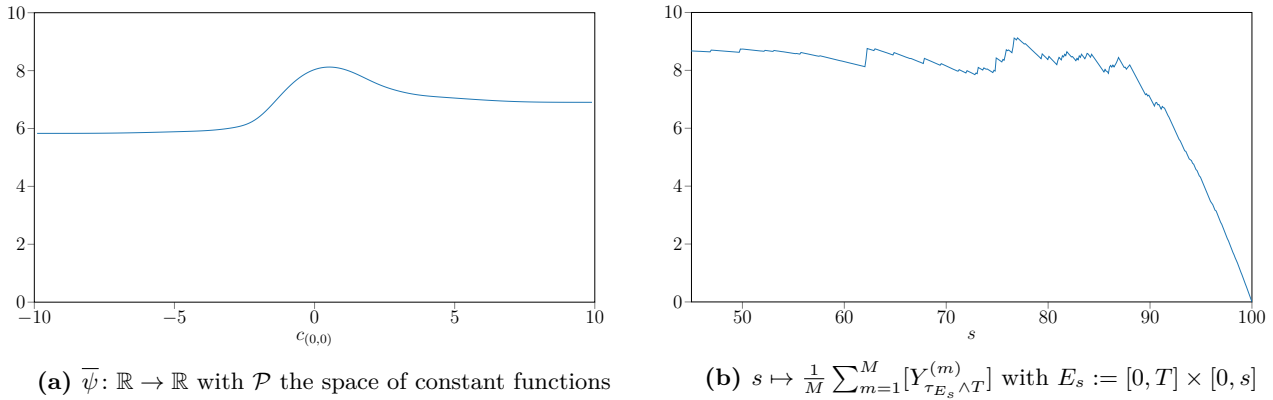}}
		\caption{$s\mapsto \frac{1}{M}\sum_{m=1}^{M}[Y^{(m)}_{\tau_{E_s}\wedge T}]$ with $E_{s}:=[0,T]\times [0,s]$}
		\label{fig:psi1theirs}
	\end{subfigure}
	\caption{Functions to be maximized in a one-parameter optimization of a randomized exercise strategy (a) and a one-parameter optimization of a deterministic strategy (b) for a one-dimensional American put option with $K=s_0=100$ and $T=1$ in the Black--Scholes model with $r=0.05$ and $\sigma=0.3$. Both plots were generated using $M=100$ sample paths with $N=100$ time steps.}
	\label{fig:psi1}
\end{figure}

 The advantage of exercise rate optimization over exercise region optimization is illustrated by \Cref{fig:psi1}. Even a simple gradient ascent algorithm could be used to maximize $\overline{\psi}$ in \Cref{fig:psi1ours}, where we show the dependence on the coefficient $c_{(0,0)}$ of the constant polynomial $p_{(0,0)}\equiv1$ for a one-dimensional put option. For comparison, this is not possible for the function shown in \Cref{fig:psi1theirs}, which arises from the optimization of deterministic exercise regions and requires the use of finite-difference stochastic-gradient algorithms.

Differentiability of $\phi$, $\psi$, and $\overline{\psi}$ with respect to $f$ is easy to show. Using the fact that $\lambda_t U_t\d t=-\d  U_t$, we obtain the simple gradient formula
\begin{align*}
\langle \nabla_{f}\phi(f,(S_{t})_{t\in\T}),h\rangle &=  -\int_{0}^{T}Y_t\d\langle \nabla_f U_t,h\rangle+\langle \nabla_f U_{T},h\rangle Y_T,\quad h\colon\T\times\R_{+}^{d}\to\R,
\end{align*}
where
\begin{align*}
\langle \nabla_f U_{t},h\rangle&=-U_t\int_{0}^{t}{h(u,{S}_u)}\d u, \quad t\in \T.
\end{align*}

\begin{figure}[h!]
	\begin{subfigure}{0.49\textwidth}
    \includegraphics[scale=0.5]{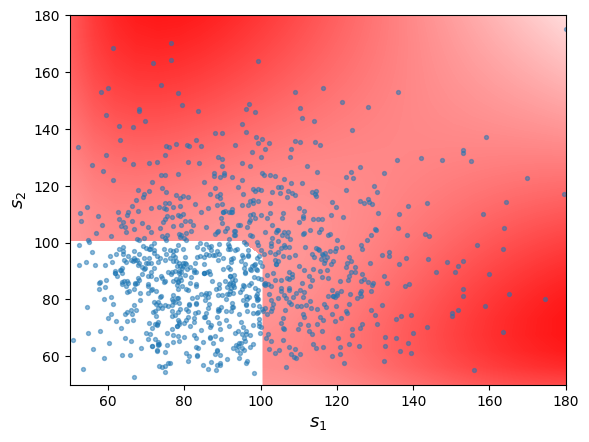}
    \caption{10th iteration}
\label{fig:ani0}
	\end{subfigure}
	\begin{subfigure}{0.49\textwidth}
    \includegraphics[scale=0.5]{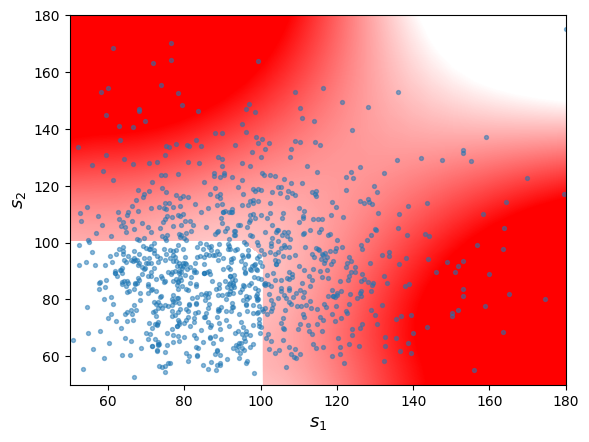}
        \caption{20th iteration}
\label{fig:ani5}
	\end{subfigure}
    \\
	\begin{subfigure}{0.49\textwidth}
    \includegraphics[scale=0.5]{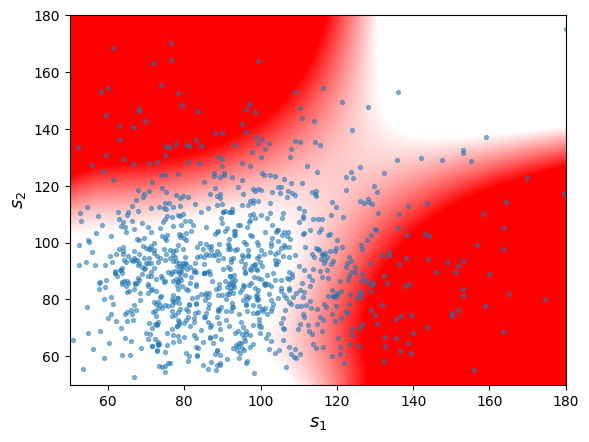}
        \caption{30th iteration}
\label{fig:ani10}
	\end{subfigure}
	\begin{subfigure}{0.49\textwidth}
    \includegraphics[scale=0.5]{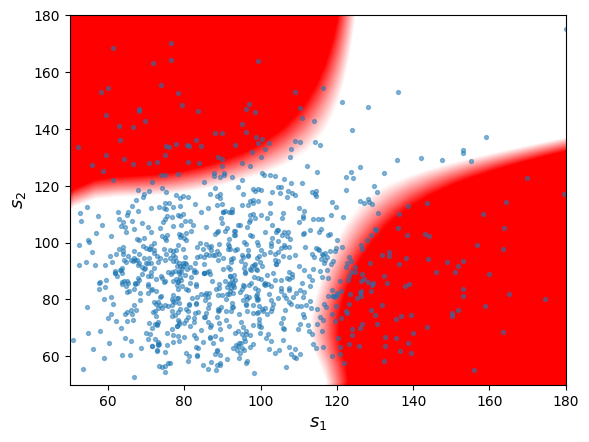}
        \caption{40th iteration}
\label{fig:ani15}
	\end{subfigure}
    \caption{Four iterations of the exercise rate optimization for a max call option (all figures show a slice of the exercise rate at $t=T/2$).
    High color intensities represent high exercise rates. The white region in the bottom left contains the points with zero payoff, $\{\payoff=0\}$. Random sample values of the two underlying securities at $T/2$ are shown in blue.}
    \label{fig:anim}
\end{figure}

\Cref{fig:anim} shows four snapshots of the search for an optimal exercise rate for max call options on two underlying securities. 

\paragraph{Accuracy}
To obtain accurate results, we must choose large enough values for the number of samples, $M$, the number of time steps, $N$, the number of iterations of the optimization routine, $\ell$, and the polynomial degree, $k$.

For a fixed exercise rate and a fixed number of time steps, convergence with respect to the number of sample paths, $M$, occurs asymptotically at the Monte Carlo rate $M^{-1/2}$. Pre-asymptotically, the number of Monte Carlo samples has to be larger than a threshold depending on the dimension of the polynomial subspace to avoid overfitting, see the next paragraph.

For a fixed, smooth exercise rate, the expected payoff converges at the weak convergence rate of the discretization scheme with respect to the number of time steps (e.g., $N^{-1}$ for the Euler--Maruyama scheme). In the limit of increasingly steep exercise rates approaching the optimal deterministic exercise regions, the weak convergence rate is expected to deteriorate to $N^{-1/2}$. However, this effect does not become noticeable in our numerical experiments (see \Cref{sec:practice0}). 

With everything else held fixed, we expect exponential or faster convergence with respect to $\ell$, depending on what type of deterministic optimization routine is used. \Cref{fig:conviter} in \Cref{sec:practice} provides numerical evidence of exponential convergence using the L-BFGS-B algorithm. 

To characterize the convergence of the optimal exercise rate with respect to $k$ under the simplifying assumptions $M=\infty$ and $N=\infty$, we note that for any polynomial $0\neq p_k\in \mathcal{P}_{k}$ the randomized exercise
strategies with exercise rates $f_{L}:=\exp(L p_k)\in
F_{k}$ converge to a deterministic strategy with early exercise region
$E_k:=\{p_k\geq 0\}$ as $L\to \infty$. Therefore, it suffices to study the approximability of the optimal exercise region $E_*$ by polynomial superlevel sets, and
the sensitivity of the expected payoff on the right-hand side of \Cref{fund} with respect to perturbations of the exercise region.  Regarding the approximability of $E_*$, we observe that if $E_*$ is a bounded $C^m$-submanifold, $m\geq 2$, of $(0,T)\times \{\payoff>0\}$, then there exists a sequence of polynomials $p_k$ such that the boundaries $B_k:=\partial E_k$ of the corresponding exercise regions $E_k:=\{p_k\geq 0\}$ satisfy 
\begin{equation}
\label{perturb}
B_k=\{(t,s)+\Theta(t,s) : (t,s) \in B_{*}\}
\end{equation}
for some $\Theta\colon B_{*}\to\R^{1+d}$ such that 
\begin{equation*}
\sup_{(t,s)\in B_{*}}|\Theta(t,s)|<C k^{-m}.
\end{equation*}
This follows from a combination of the multi-dimensional Jackson theorem \cite{BagbyBosLevenberg2002} with a partition of unity and elementary geometry. Regarding the sensitivity of the expected payoff, \cite{gobet2006sensitivities} showed differentiability with respect to perturbations of the exercise region in spatial directions under the assumption that $(0,s_0)\not\in E_*$ and that the payoff function lies in some Hölder space $C^{1,\alpha}$, $\alpha>0$. Unfortunately, this result is not quite general enough for our purposes, since we require bounds with respect to general, spatio-temporal perturbations of the domain (as in \Cref{perturb}) and for payoff functions that are only Lipschitz.

A rigorous analysis of the interplay of the various discretizations will be the topic of future work;
some numerical results are presented in \Cref{sec:practice0} below.

\paragraph{Overfitting}
Choosing a subspace with a large number of degrees of freedom, $B\gg 1$, to improve the flexibility of the candidate exercise rates increases the cost of computations and the risk of \emph{overfitting}. This means that the value of $\overline{\psi}(\bm{c}^*)$ at the optimized coefficients $\bm{c}^{*}$ may overestimate the true value $\psi(f_{\bm{c}^*})$ unless a correspondingly large number $M=M(B)$ of sample paths is used. Numerical experiments indicate that $M(B)\approx CB^2$ for some $C>0$ but we were not able to prove such a formula. In practice, we can simply compute an unbiased estimate of $\psi(f_{\bm{c}^{*}})$ using a new set of sample paths $(\tilde{S}^{(m)}_{t})_{t\in\T}$, $1\leq m\leq {M}$; %According to \Cref{equality},  test value is a lower bound of the option price.
 similar techniques are used in classical regression-based methods such as the
 Longstaff--Schwartz algorithm. Following statistical learning terminology, we
 refer to the biased and unbiased estimators of $\psi(f_{\bm{c}^{*}})$ as
 \emph{training} and \emph{test} values, respectively. One way to avoid
 overfitting is to recompute the test value at each step of the optimization
 and to terminate as soon as the test value decreases. Note that, as in the
 case of the Longstaff--Schwartz algorithm, the test values are biased low,
 i.e., are Monte Carlo estimates of \emph{lower bounds} of the option price.

\section{Numerical experiments}
\label{sec:practice}
Throughout this section, we use the L-BFGS-B algorithm with initial coefficients $\bm{c}\equiv 0$  to maximize $\overline{\psi}$. 
\tikzset{fontscale/.style = {font={\small #1}}
}
\subsection{Convergence with respect to discretization parameters}
\label{sec:practice0}
\begin{figure}
	\centering
	\begin{subfigure}{0.48\textwidth}
	% This file was created by matplotlib2tikz v0.6.17.
\begin{tikzpicture}

\definecolor{color0}{rgb}{0.12156862745098,0.466666666666667,0.705882352941177}
\definecolor{color1}{rgb}{1,0.498039215686275,0.0549019607843137}
\definecolor{color2}{rgb}{0.172549019607843,0.627450980392157,0.172549019607843}
\begin{axis}[
xlabel={$n$},
xmin=-0.35, xmax=7.35,
ymin=8.87716757040387, ymax=10.4141372641039,
tick align=outside,
tick pos=left,
xmajorgrids,
xminorgrids,
x grid style={white!50.19607843137255!black},
ymajorgrids,
yminorgrids,
y grid style={white!50.19607843137255!black},
legend entries={{Test value},{Training value},{Reference value}},
legend style={draw=white!80.0!black},
legend cell align={left}
]
%\addlegendimage{no markers, color0}
%\addlegendimage{no markers, color1}
%\addlegendimage{no markers, white!50.19607843137255!black}
\addplot [semithick, color0, mark=*, mark size=1.5, mark options={solid}]
table {%
0 10.3442750052993
1 9.54882601722182
2 10.0311534054694
3 9.85588617579995
4 9.81794806472764
5 9.86670489662966
6 9.85865544976034
7 9.86763717617984
};
\addplot [semithick, color2, mark=*, mark size=1.5, mark options={solid}]
table {%
0 10.0835004743148
1 8.94702982920841
2 10.1186000219387
3 9.82548111236231
4 9.87142406832498
5 9.81682559451678
6 9.87149886932064
7 9.86799472386241
};
\addplot [line width = 1.5, color1, solid]
table {%
0 9.8701
1 9.8701
2 9.8701
3 9.8701
4 9.8701
5 9.8701
6 9.8701
7 9.8701
};
\end{axis}

\end{tikzpicture}
	\caption{$k=2$}
	\label{fig:conv1d_k2_0}
	\end{subfigure}
\hfill
	\begin{subfigure}{0.48\textwidth}
		\vspace*{-0.0em}
	% This file was created by matplotlib2tikz v0.6.17.
\begin{tikzpicture}

\definecolor{color0}{rgb}{0.12156862745098,0.466666666666667,0.705882352941177}

\begin{axis}[
xlabel={$n$},
xmin=-0.35, xmax=7.35,
ymin=0.000191352205237974, ymax=0.0657592018253286,
ymode=log,
every y tick label/.style = {
rotate=0
},
xtick pos=left,
ytick pos=both,
x grid style={white!69.01960784313725!black},
ymajorgrids,
yminorgrids,
y grid style={white!69.01960784313725!black},
legend style={at={(0.03,0.03)}, anchor=south west, draw=white!80.0!black},
legend cell align={left}
]
%\addlegendimage{no markers, color0}
\addplot [line width = 1.5, color0, dashed]
table {%
0.368421052631579 0.0504287510388456
3.68421052631579 0.00506437809079436
7 0.000508597276715443
};
\addplot [semithick, color0, mark=*, mark size=1.5, mark options={solid}]
table {%
0 0.0480415603995224
1 0.0325502257097884
2 0.0163173023038675
3 0.00144008917843341
4 0.0052838304852392
5 0.00034397861929879
6 0.00115951715176774
7 0.000249523694811968
};
%\addlegendentry{}
\addlegendentry{$2^{-n}$}
\end{axis}

\end{tikzpicture}
	\caption{Relative error of test value, $k=2$}
	\label{fig:conv1d_k2_1}
	\end{subfigure}
\vskip\baselineskip
 	\begin{subfigure}{0.48\textwidth}
	% This file was created by matplotlib2tikz v0.6.17.
\begin{tikzpicture}

\definecolor{color0}{rgb}{0.12156862745098,0.466666666666667,0.705882352941177}
\definecolor{color1}{rgb}{1,0.498039215686275,0.0549019607843137}
\definecolor{color2}{rgb}{0.172549019607843,0.627450980392157,0.172549019607843}
\begin{axis}[
xlabel={$n$},
xmin=-0.3, xmax=6.3,
ymin=6.75329829256425, ymax=10.1959626716539,
tick align=outside,
tick pos=left,
xmajorgrids,
xminorgrids,
x grid style={white!50.19607843137255!black},
ymajorgrids,
yminorgrids,
y grid style={white!50.19607843137255!black},
legend style={at={(0.97,0.03)}, anchor=south east, draw=white!80.0!black},
legend entries={{Test value},{Training value},{Reference value}},
legend cell align={left}
]
%\addlegendimage{no markers, color0}
%\addlegendimage{no markers, color1}
%\addlegendimage{no markers, white!50.19607843137255!black}
\addplot [semithick, color0, mark=*, mark size=1.5, mark options={solid}]
table {%
0 8.71382942568109
1 8.85167902876452
2 9.39605327615948
3 9.82037810380882
4 9.82784439119477
5 9.82320862089329
6 9.84216000934811
};
\addplot [semithick, color2, mark=*, mark size=1.5, mark options={solid}]
table {%
0 6.90978303706832
1 9.68763178172029
2 9.41069176886661
3 10.0394779271498
4 9.85268547077696
5 9.864304232928
6 9.86880425272541
};
\addplot [line width = 1.5, color1, solid]
table {%
0 9.8701
1 9.8701
2 9.8701
3 9.8701
4 9.8701
5 9.8701
6 9.8701
};
\end{axis}

\end{tikzpicture}
	\caption{ $k=1$}
	\label{fig:conv1d_k0}
\end{subfigure}
\begin{subfigure}{0.48\textwidth}
	\vspace*{-0.0em}
	% This file was created by matplotlib2tikz v0.6.17.
\begin{tikzpicture}

\definecolor{color0}{rgb}{0.12156862745098,0.466666666666667,0.705882352941177}
\definecolor{color1}{rgb}{1,0.498039215686275,0.0549019607843137}
\definecolor{color2}{rgb}{0.172549019607843,0.627450980392157,0.172549019607843}
\begin{axis}[
xlabel={$n$},
xmin=-0.35, xmax=7.35,
ymin=9.03342899939514, ymax=10.0932081072973,
tick align=outside,
tick pos=left,
xmajorgrids,
xminorgrids,
x grid style={white!50.19607843137255!black},
ymajorgrids,
yminorgrids,
y grid style={white!50.19607843137255!black},
legend style={at={(0.97,0.03)}, anchor=south east, draw=white!80.0!black},
legend entries={{Test value},{Training value},{Reference value}},
legend cell align={left}
]
%\addlegendimage{no markers, color0}
%\addlegendimage{no markers, color1}
%\addlegendimage{no markers, white!50.19607843137255!black}
\addplot [semithick, color0, mark=*, mark size=1.5, mark options={solid}]
table {%
0 9.25466365981809
1 10.0450363296654
2 9.08160077702706
3 9.26366605182998
4 9.35304319131022
5 9.36465535302211
6 9.37108412677606
7 9.34477175470685
};
\addplot [semithick, color2, mark=*, mark size=1.5, mark options={solid}]
table {%
0 9.09485004399888
1 9.46734979567127
2 9.67522024602217
3 9.23787434513326
4 9.34082409817102
5 9.34857681487816
6 9.33902438531038
7 9.36491149228061
};
\addplot [line width = 1.5, color1, solid]
table {%
0 9.8701
1 9.8701
2 9.8701
3 9.8701
4 9.8701
5 9.8701
6 9.8701
7 9.8701
};
\end{axis}

\end{tikzpicture}
	\caption{$k=0$}
	\label{fig:conv1d_k1}
\end{subfigure}
\caption{Exercise rate optimization with polynomial degree $0\leq k\leq 2$, $M_n=200\times 4^{n}$ and $N_n=2^n$, $0\leq n\leq 7$ applied to a one-dimensional American put option in the Black--Scholes model with $\sigma=0.3$, $r=0.05$, $K=100$, $s_0=100$, and $T=1$.}
\label{fig:conv1d_k2}
\end{figure}
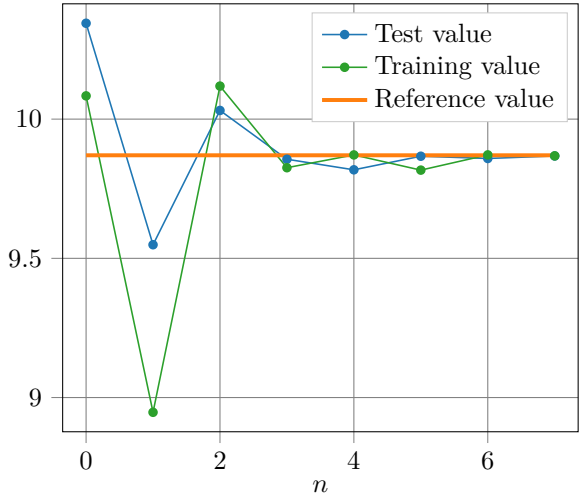
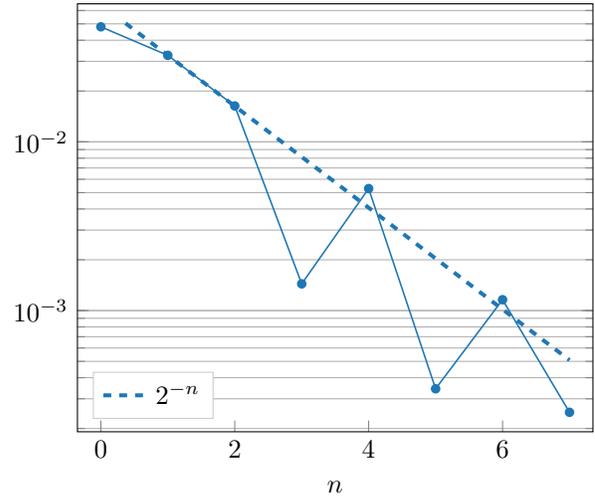
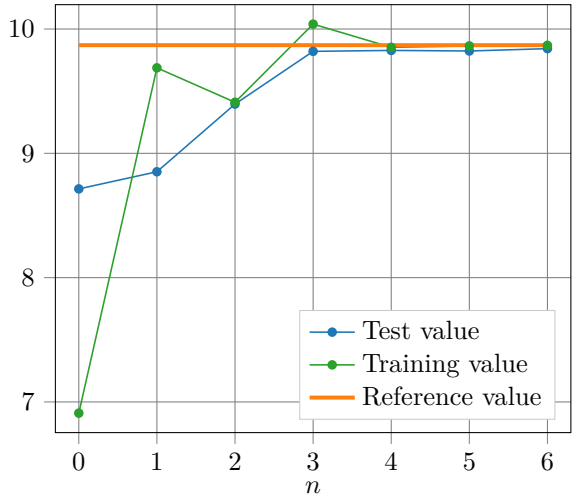
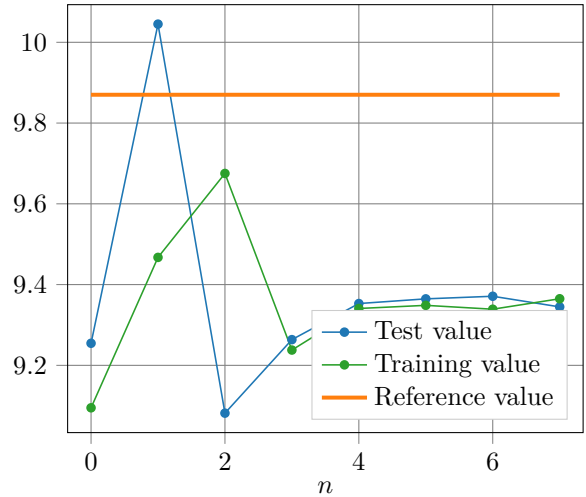

In this subsection, we study the convergence of our method with respect to the discretization parameters $M$, $N$, $k$, and $\ell$ by pricing the vanilla put option from \Cref{fig:psi1} with strike $K=100$ and expiry $T=1$ in the Black--Scholes model with volatility $\sigma=0.3$, risk-free interest rate $r=0.05$, and spot price $s_0=100$. Using a binomial tree algorithm with $\num{50000}$ levels (i.e., $\num{50000}$ time steps and $\num{50000}$ spatial discretization nodes at $T=1$), we obtain the reference value $V^*=9.8701$. %exact result: .870129; with PremiaFDexplicit(10^4 steps): .870064 PremiaEuler (10^5): .8701
 \Cref{fig:conv1d_k2_0,fig:conv1d_k2_1} show that the prices found through exercise rate optimization with polynomial degree $k=2$ and  $M_n:=200\times 4^{n}$ sample paths with $N_n:=2^{n}$ time-steps converge towards this reference value as $n\to\infty$. 
In particular, our maximization does not get stuck in local optima of $\overline{\psi}$. Furthermore, \Cref{fig:conv1d_k2_0} shows that test and training values converge at roughly the same speed, which means that we do not suffer from overfitting. This is not surprising, since the space of bivariate quadratic polynomials is only 6-dimensional. We restrict the following plots to the test value, which constitutes an unbiased estimate of the quality of a given exercise rate.

In the logarithmic scale of  \Cref{fig:conv1d_k2_1}, we see that our approximations converge to the reference value at roughly the rate $2^{-n}=\mathcal{O}(N_n^{-1}+M_n^{-1/2})$.
 We obtain an accuracy of about four significant digits, despite using only quadratic polynomials for the exercise boundary approximation. This confirms that singularities of the exercise boundary as a function of time do not pose a problem for our polynomial approximation scheme. For comparison, \Cref{fig:conv1d_k1,fig:conv1d_k0} show results for $k\in \{0,1\}$, that is, for constant exercise rates and for exercise rates that depend only linearly on space and time, respectively. For $k=1$, the results are astoundingly similar to the case $k=2$, though closer inspection on a logarithmic scale reveals stagnation at a relative error of $0.5\%$. For $k=0$, our method stagnates around the value $9.35$, which is roughly the price of a European option with the same parameters. 

\begin{figure}[h!]
 	\centering
 		% This file was created by matplotlib2tikz v0.6.17.
\begin{tikzpicture}

\definecolor{color0}{rgb}{0.12156862745098,0.466666666666667,0.705882352941177}

\begin{axis}[
xlabel={$\ell$},
xmin=-0.35, xmax=40.35,
%ymin=1e-9, ymax=0.3,
ymode=log,
%tick align=outside,
%every y tick label/.style = {%
%	rotate=0
%},
%ylabel={rel. error},
%ytick style={draw=none},
xtick style={draw=none},
%xtick pos=bottom,
%xticks off
%ytick pos=none,
%x grid style={white!69.01960784313725!black},
ymajorgrids,
yminorgrids,
xmajorgrids,
max space between ticks=20,
%y grid style={white!69.01960784313725!black},
%legend style={at={(0.03,0.03)}, anchor=south west, draw=white!80.0!black},
%legend cell align={left}
]
%\addlegendimage{no markers, color0}
%\addplot [line width = 1.5, color0, dashed]
%table {%%
%	0.368421052631579 0.0504287510388456
%	3.68421052631579 0.00506437809079436
%	7 0.000508597276715443
%};
\addplot [semithick, color0, mark=*, mark size=1.5, mark options={solid}]
table[y expr=-\thisrowno{1}/0.09838, y index=1]{%
0 -0.012668637187999723
1 -0.006752218602199184
2 -0.004555238391842548
3 -0.002722966056002321
4 -0.0017813036348595518
5 -0.0010277581063193875
6 -0.0006253228945643091
7 -0.0004726653646820733
8 -0.00036113031150460106
9 -0.0002976771668733863
10 -0.00024695986060356556
11 -0.00021387811299035386
12 -0.00019239255549841072
13 -0.00017061763412111464
14 -0.00014917584427431851
15 -0.00012312576965989308
16 -0.00011366502630927311
17 -9.795526975267532e-05
18 -8.934310877142804e-05
19 -7.868678422474906e-05
20 -6.139990311973542e-05
21 -4.797122818663013e-05
22 -3.995198586656534e-05
23 -3.085957461056166e-05
24 -2.6051452476527626e-05
25 -2.1982895393610202e-05
26 -1.8771963246982937e-05
27 -1.5632911162036245e-05
28 -1.1643098628777437e-05
29 -6.384675888718161e-06
30 -5.160903443324205e-06
31 -3.9348080226181414e-06
32 -3.1978046333563936e-06
33 -2.4284257945789145e-06
34 -1.6202210056798227e-06
35 -6.636394501990939e-07
36 -3.3693767141629305e-07
37 -1.4799899167305952e-07
38 -7.798086876231736e-08
39 -3.9642743226986354e-08
40 -1.9410056728563774e-08
41 -9.159732000663112e-09
42 -5.253393872695078e-09
43 -2.4869872411459326e-09
44 -1.2683161326743075e-09
45 -6.280235914646326e-10
46 -3.144223909012922e-10
47 -1.6296654303804559e-10
48 -5.587295903719536e-11
49 -3.274577831113845e-11
50 -1.5342824233322006e-11
51 -7.857339778816197e-12
52 -3.888486754810572e-12
53 -1.949856942573547e-12
54 -9.726941474497153e-13
55 -4.86166662483356e-13
56 -2.427780199099061e-13
57 -1.2126410986468272e-13
58 -6.053491041768666e-14
59 -3.019806626980426e-14
60 -1.504352198367087e-14
61 -7.466249840604178e-15
62 -3.6637359812630166e-15
63 -1.8041124150158794e-15
64 -8.881784197001252e-16
65 -4.440892098500626e-16
66 -2.220446049250313e-16
67 -1.1102230246251565e-16
68 -5.551115123125783e-17
69 -2.7755575615628914e-17
70 -1.3877787807814457e-17
};
%\addlegendentry{}
\end{axis}

\end{tikzpicture}
 	\caption{Convergence with respect to the number of function evaluations, $\ell$, in the training step of exercise rate optimization for an American put option using the L-BFGS-B algorithm.}
 	\label{fig:conviter}
 \end{figure}
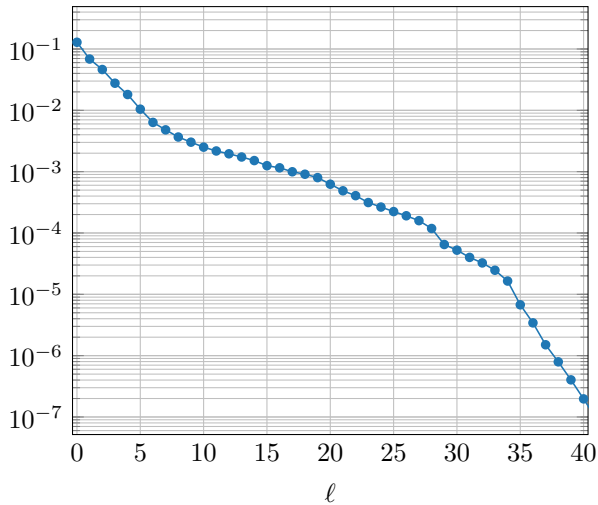

To study the effects of $M$, $N$, and $k$, we performed experiments in this and the following subsection with the tolerance of the L-BFGS-B optimization set to machine precision, which required between $70$ and $200$ function evaluations to achieve.
However, an error comparable to that of the remaining discretization errors can already be achieved with significantly fewer evaluations. Indeed, for $n=4$ and $k=2$ the relative error between $\overline{\psi}(\bm{c}_{\ell})$ and the final value is already below $0.1\%$ when $\ell=20$ (\Cref{fig:conviter}). 
For this reason, we limit the number of iterations below to $20$.
%In the remaining subsections, %we employ the early termination strategy discussed in the last paragraph of the previous section, which usually terminates after 15-25 steps. 

\subsection{Comparison with Longstaff--Schwartz algorithm}%l=200
\label{sec:practice1}
In this subsection, we consider \emph{basket put options} on linear combinations of $d\in\{2,5\}$ underlying assets. The payoff function of such options is given by $\payoff (s):=(K-c\cdot {s})^{+}$ for $K>0$ and $c\in\R^{d}$. In our experiments, we use $K:=100$ and $c_i:=1/d$, $1\leq i\leq d$.

We compare our method to the Longstaff--Schwartz algorithm, as implemented in the freely available version 16 of the derivative pricing software  Premia\footnote{\url{https://www.rocq.inria.fr/mathfi/Premia}}.
Like our method, the Longstaff--Schwartz algorithm requires specification of the number of sample paths, the number of time-steps used for their simulation, and the polynomial degree, which controls the accuracy of approximations of the value function. For simplicity, we restrict the simulations in this section to $N=8$ time steps. To prevent our comparison being skewed by the fact that the two algorithms use different sample paths, we use the same large number of $M=3.2\times 10^6$ samples for both. Finally, we use a risk-free interest rate $r=0.05$ and a diagonal volatility matrix $\Sigma_{ij}=0.3^2\delta_{ij}$, $1\leq i,j\leq d$ in the underlying Black--Scholes model with $s_0=(100,\dots,100)$.

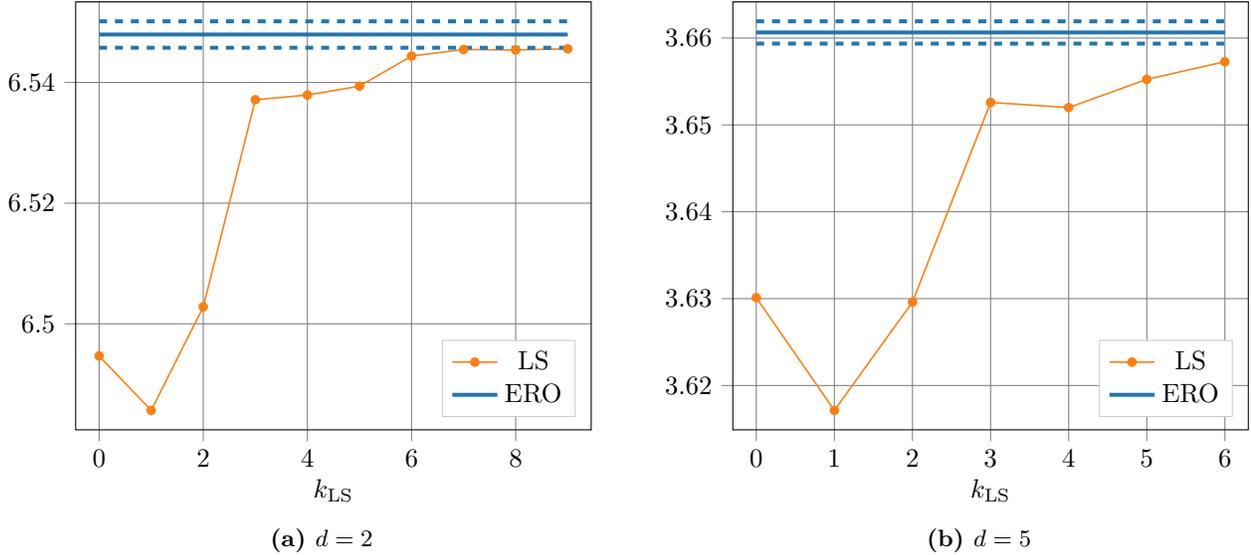
\begin{figure}[h!]
	\centering%
	\begin{subfigure}{0.49\textwidth}
	% This file was created by matplotlib2tikz v0.6.17.
\begin{tikzpicture}

\definecolor{color0}{rgb}{0.12156862745098,0.466666666666667,0.705882352941177}
\definecolor{color1}{rgb}{1,0.498039215686275,0.0549019607843137}
\definecolor{color2}{rgb}{0.172549019607843,0.627450980392157,0.172549019607843}

\begin{axis}[
xlabel={$k_{\mathrm{LS}}$},
xmin=-0.45, xmax=9.45,
ymin=6.48243327132545, ymax=6.55335530216547,
tick align=outside,
tick pos=left,
xmajorgrids,
xminorgrids,
x grid style={white!50.19607843137255!black},
ymajorgrids,
yminorgrids,
y grid style={white!50.19607843137255!black},
legend style={at={(0.97,0.03)}, anchor=south east, draw=white!80.0!black},
]
\addplot [semithick, color1, mark=*, mark size=1.5, mark options={solid}]
table {%
0 6.494686
1 6.485657
2 6.502813
3 6.537143
4 6.537915
5 6.539393
6 6.544376
7 6.545461
8 6.54539
9 6.545565
};
\addplot [line width = 1.5, color0,solid]
table {%
0 6.54794392040128
9 6.54794392040128
};;
\addplot [line width = 1.5, color0, dashed, forget plot]
table {%
0 6.54575626731163
9 6.54575626731163
};
\addplot [line width = 1.5, color0, dashed, forget plot]
table {
0 6.55013157349093
9 6.55013157349093
};
\addlegendentry{LS}
\addlegendentry{ERO}
\end{axis}

\end{tikzpicture}
\caption{$d=2$}	
\end{subfigure}
\begin{subfigure}{0.49\textwidth}
	% This file was created by matplotlib2tikz v0.6.17.
\begin{tikzpicture}

\definecolor{color0}{rgb}{0.12156862745098,0.466666666666667,0.705882352941177}
\definecolor{color1}{rgb}{1,0.498039215686275,0.0549019607843137}
\definecolor{color2}{rgb}{0.172549019607843,0.627450980392157,0.172549019607843}

\begin{axis}[
xlabel={$k_{\mathrm{LS}}$},
xmin=-0.3, xmax=6.3,
ymin=3.61489649413897, ymax=3.66416562308172,
tick align=outside,
tick pos=left,
xmajorgrids,
xminorgrids,
x grid style={white!50.19607843137255!black},
ymajorgrids,
yminorgrids,
y grid style={white!50.19607843137255!black},
legend style={at={(0.97,0.03)}, anchor=south east, draw=white!80.0!black},
]
\addplot [semithick, color1, mark=*, mark size=1.5, mark options={solid}]
table {%
0 3.630112
1 3.617136
2 3.629618
3 3.652589
4 3.652003
5 3.655242
6 3.657282
};
\addplot [line width = 1.5, color0]
table {%
0 3.66064268342768
6 3.66064268342768
};
\addplot [line width = 1.5, color0, dashed, forget plot]
table {%
0 3.65935924963468
6 3.65935924963468
};
\addplot [line width = 1.5, color0, dashed, forget plot]
table {%
0 3.66192611722069
6 3.66192611722069
};
\addlegendentry{LS}
\addlegendentry{ERO}
\end{axis}

\end{tikzpicture}
\caption{$d=5$}	
\end{subfigure}
	\caption{Convergence of the Longstaff--Schwartz algorithm (LS) for $\{2,5\}$-dimensional basket put options with increasing polynomial degree $k_{\mathrm{LS}}$ to reference values computed via exercise rate optimization (ERO) with polynomial degree $k_{\mathrm{ERO}}=2$ and $95\%$ confidence bands (dashed).}
	\label{fig:nd}
\end{figure}

To emphasize the efficiency of exercise rate optimization with respect to the polynomial degree, we compute reference values $V^*=6.5479$ and $V^*=3.6606$ using exercise rate optimization with polynomial degree $k_{\mathrm{ERO}}=2$ for $d=2$ and $d=5$, respectively. \Cref{fig:nd} shows that the Longstaff--Schwartz algorithm converges to these values as $k_{\mathrm{LS}}\to\infty$, but only achieves a comparable performance for $k\approx 6$. We show $95\%$ confidence bands around our reference value, which are based on the empirical variance in the evaluation of our test value. From these we see that the remaining difference between the two methods can be explained by the random sampling error.

\paragraph{Runtime comparison} 

To obtain a fair runtime comparison, we created a Python package\footnote{\url{https://pypi.org/project/pryce/}} with straightforward implementations of both algorithms, which we ran on a 12 core Intel Xeon X5650 CPU.

For the same polynomial degree, exercise rate optimization is slower than the Longstaff--Schwartz algorithm. However, as we have seen above, the latter requires larger polynomial degrees for accurate results. Since the ratio between the dimensions of polynomial subspaces with degrees $k=2$ and $k>2$ grows with respect to the dimension of the domain, exercise rate optimization returns accurate results faster than the Longstaff--Schwartz algorithm in high-dimensional examples.

For example, for a basket put option as above with $d=10$, the Longstaff--Schwartz algorithm returns $2.235$ with $k=2$ after $530$ seconds and $2.237$ with $k=4$ after $7437$ seconds. 
Exercise rate optimization, on the other hand, returns $2.240$ with $k=2$ after $2493$ seconds. All these results were obtained with the same $3.2\times 10^6$ Brownian motion samples.

\subsection{Max call options}%l=20
\begin{figure}[h!]
	\centering
		\input{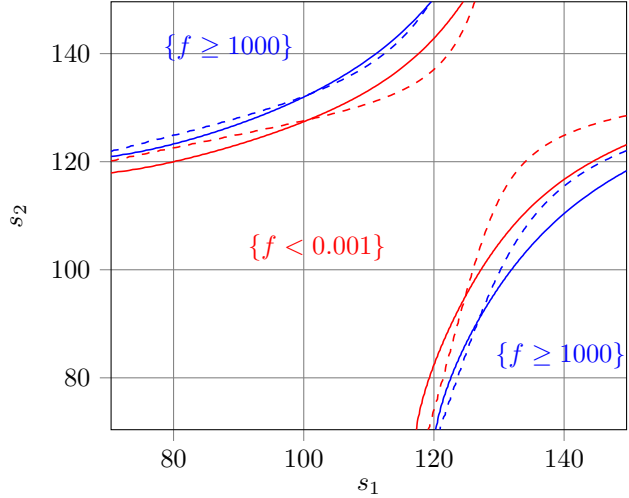}
	\caption{Level sets of optimal exercise rates for a max call option with $k=2$ (dashed) and $k=3$ (solid).}
	\label{fig:exratemaxcall}
\end{figure}

\label{sec:practice1b}
In this subsection, we consider max call options on two underlying assets, for which $g(s):=\max\{(s_1-K)^{+},(s_2-K)^{+}\}$. These max call options present an interesting challenge for our method, since the optimal exercise region  at any time before expiry has two connected components \cite{broadie1997valuation}. Lower and upper bounds for the option prices in the Black--Scholes model with $r=0.05$, $\Sigma_{ij}=0.2^2\delta_{ij}$, $K=100$, $N=8$ and dividend $\delta=0.1$ are taken from \cite{andersen2004primal} and provided in \Cref{table:maxcall} alongside the results of our method for $k\in\{1,2,3\}$ and $M=\SI{1000000}{}$.%
\renewcommand{\arraystretch}{1.1}
\begin{table}[h!]
	\centering
\begin{tabular}{cc|cccc}
&	& &  \multicolumn{3}{c}{\raisebox{-0.5em}{$k$}}\\
& & \raisebox{0.1em}{95\% CI} & $1$ & $2$ & $3$% & $4$ %& $5$ & $6$
\\ \hline
&90 & [8.053,8.082] %
 & 7.126% 7.126277940654041,
 & 8.009%8.008640458361191, 
 & 8.039%8.038699326950467,
\\
$s_0$ &100 & [13.892,13.934] %
 & 12.311%12.310561361714997 
 & 13.821%13.821168023674743, 
 & 13.865%13.865020396083663,
 \\
&110 & [21.316, 21.359] %
 & 19.133%19.13334969635388 
 & 21.220%21.220316924559907, 
 &21.256%21.256491128521274,
 \\
\end{tabular}
\caption{Prices of max call option. 95\% confidence intervals (CI) taken from \cite{andersen2004primal}.}
\label{table:maxcall}
\end{table}
The optimized exercise rates with $k\in\{2,3\}$ are shown in \Cref{fig:exratemaxcall}.
As expected, they are almost deterministic, which means that they exhibit steep slopes from values close to zero to values close to infinity. Since the specific values are irrelevant, we restrict our plots to the level sets of exercise rate $0.001$ and $1000$. The results in this subsection were obtained using a maximal number of $20$ optimization steps. Performing more steps would further reduce the distance between these level sets without a noticeable difference in the resulting option price. 
As predicted by theory, there are two disjoint regions of high exercise rates. Furthermore, due to the symmetry of the underlying model and the payoff,  the optimized exercise rate is almost axisymmetric even though we do not enforce this symmetry. While modeling the disconnected regions is not possible with log-linear exercise rates available for $k=1$, the hyperbolic conic sections available with $k=2$ already provide satisfactory approximations. 

\subsection{Stochastic volatility}%l=20
\label{sec:practice2}
In this subsection, we apply our method to pricing in a stochastic volatility model. 

For this purpose, we consider the basic Heston model as described in \cite{heston1993closed}, which models the evolution of a single underlying asset $X_t$ and its instantaneous variance $\variance_t$ using the coupled system of stochastic differential equations
\begin{align}
\label{hestonunderlying}
\d X_t &= \mu X_t\d t +\sqrt{\variance_t} X_t \d W^{X}_t,\\
\label{hestonvolatility}
\d\variance_t &= \kappa(\theta-\variance_t)\d t+\xi \sqrt{\variance_t}\d W^{\variance}_{t},
\end{align}
where $\mu>0, \kappa>0$, $\theta>0$, $\xi>0$ with $2\kappa\theta>\xi^2$, and $W^{X}_t$ and $W^{\variance}_t$ are Wiener processes with correlation $-1\leq\rho\leq 1$.

Since our method requires Markovian markets, we must include the volatility and define $S_t:=(X_t,\variance_t)$, $t\in\T$. This means that knowledge of the current volatility is required to make optimal exercise decisions in stochastic volatility models.

To obtain a risk neutral measure, we replace $\mu$ with the risk-free rate $r=0.05$ in \Cref{hestonunderlying}. We choose the remaining parameters $\kappa=3$, $\theta=0.05$, $\xi=0.5$, $\rho=-0.5$ and compute estimates of $v_K(s_0)$ for a put option with $s_0=(100,0.15)$ and $25$ different values of the strike $K\in[90,150]$. For this purpose, we use polynomials of degree $k\in\{0,1,2\}$ and $M=\SI{100000}{}$ samples with $N=32$ time steps.

\begin{figure}[h!]
	\centering
	% This file was created by matplotlib2tikz v0.6.17.
\begin{tikzpicture}

\definecolor{color0}{rgb}{0.12156862745098,0.466666666666667,0.705882352941177}
\definecolor{color1}{rgb}{1,0.498039215686275,0.0549019607843137}
\definecolor{color2}{rgb}{0.172549019607843,0.627450980392157,0.172549019607843}
\begin{axis}[
xmin=77, xmax=143,
ymin=-1.33, ymax=43.23189145,
tick align=outside,
tick pos=left,
xlabel={$K$},
x grid style={white!69.01960784313725!black},
y grid style={white!69.01960784313725!black},
xmajorgrids,
xminorgrids,
x grid style={white!50.19607843137255!black},
ymajorgrids,
yminorgrids,
y grid style={white!50.19607843137255!black},
legend style={at={(0.05,0.95)}, anchor=north west, draw=white!80.0!black},
]

\addplot [semithick, color1]
table {%
70.        	1.303993
75.       	1.946491
80.       	2.808061
85.      	3.931832
90 			5.362171
95 			7.129598
100 		9.273402
105 		11.809727
110 		14.750833
115 		18.105119
120 		21.843436
125 		25.938127
130 		30.350797
135 		35.002031
140 		39.872711
145 		44.810764
150			49.780551
};

\addplot [semithick, color0]%M=100000, k=2, N=32
table {%
70.           1.30769717
75.           1.94890402
80.           2.84596171
85.           3.94040269
90.           5.36039998
95.           7.23147049
100.          9.38294894
105.          11.88617836
110.          14.88372023
115.          18.15446168
120.          21.86884886
125.          25.98771441    
130.          30.             
135.          35.             
140.          40.             
145.          45.            
150.          50.       
};

\addplot [semithick, color0,style=dashed]%M=100000, k=1, N=32
table {%
70.0 1.250232505092076
75.0 1.9014966979295094
80.0 2.722973287837494
85.0 3.9744712783888247
89.99999999999999 5.391194165889306
95.0 7.115380323398639
100.0 9.279877893528193
105.0 11.804866048289478
110.00000000000001 14.746429094281046
114.99999999999999 18.108135513082992
120.0 21.86247094959671
125.0 25.0
130.0 29.99999999999999
135.0 35.0
140.0 40.0
145.0 45.000000000000014
150.0 50.0
};

\addplot [semithick, color0,style=dotted]%M=100000, k=0, N=32
table {%
70.0 1.2719337175458927
75.0 1.8598820418296647
80.0 2.7127469203822883
85.0 3.7080713665944556
89.99999999999999 5.030238939472286
95.0 6.711699791757039
100.0 8.65584360937518
105.0 10.92989530269662
110.00000000000001 13.492232399797214
114.99999999999999 16.549368515495058
120.0 20.152696658032507
125.0 25.0
130.0 29.999999999999222
135.0 34.99999999999919
140.0 39.99999999999933
145.0 44.99999999999918
150.0 49.9999999999992
};

\addplot [semithick, gray]%European
table {%
70.0 1.2510096408508697
75.0 1.8341124433331208
80.0 2.6912084956307867
85.0 3.749277886135631
89.99999999999999 5.028799129811209
95.0 6.644613227515076
100.0 8.602246948242364
105.0 10.921841088955524
110.00000000000001 13.564419895436028
114.99999999999999 16.46997444696904
120.0 19.62882669982641
125.0 23.156575335101547
130.0 26.805580564978133
135.0 30.965554951704032
140.0 35.00452951600552
145.0 39.419990660972616
150.0 43.729061284082206
};

\addlegendentry{FD}
\addlegendentry{ERO ($k=2$)}
\addlegendentry{ERO ($k=1$)}
\addlegendentry{ERO ($k=0$)}
\addlegendentry{European}
\end{axis}

\end{tikzpicture}
	\caption{Dependence of the put option price on the strike in the Heston model; computed using exercise rate optimization (ERO) and a finite-difference method (FD).}
	\label{fig:heston1d}
\end{figure}
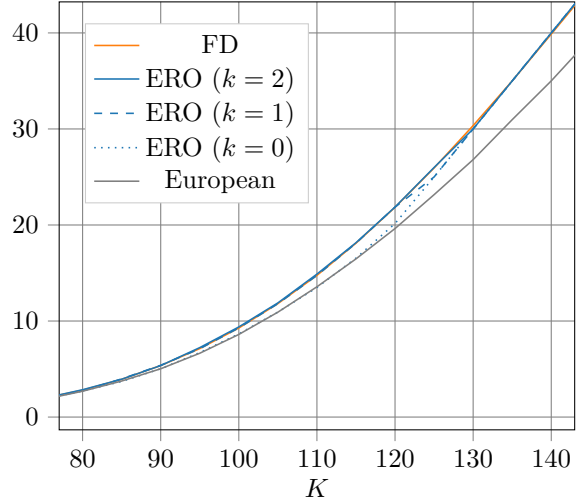

For comparison, we also show the results of the finite difference method \verb|FD_Hout_Heston| implemented in Premia, with $32$ time steps and a grid of $100\times 100$ nodes in the discretization of the stock-volatility plane. The results are shown in \Cref{fig:heston1d}.  
The maximal relative difference between the two methods is $1\%$ and occurs around $K^*=130$. Up to roundoff error, the prices computed by our method are equal to $K-100$ for all $K\geq K^*$. This behavior is expected, since for large enough $K$ the initial point $(100,0.15)$ lies within the optimal exercise region and the option is thus exercised immediately.

\begin{figure}[h!]
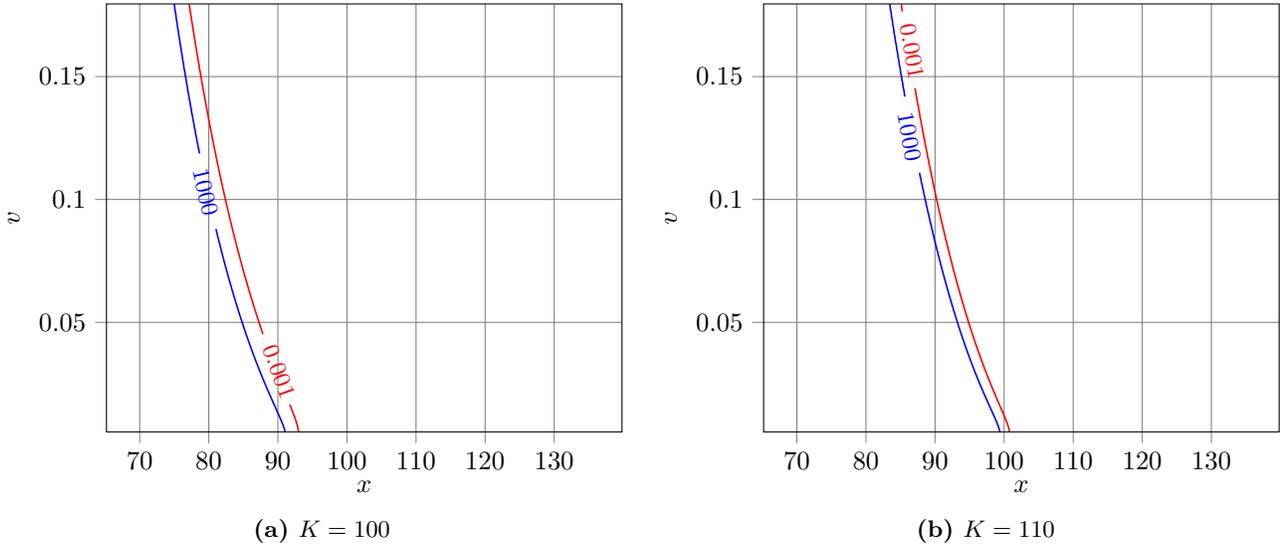

	\begin{subfigure}{0.49\textwidth}
	\input{figures/contourheston100}
	\caption{$K=100$}
	\end{subfigure}
\begin{subfigure}{0.5\textwidth}
	\input{figures/contourheston110}
	\caption{$K=110$}
\end{subfigure}
\caption{Level sets of optimal exercise rates at $t=0.5$ for a put option in the Heston model.}
\label{fig:exrate}
\end{figure}

\Cref{fig:exrate} shows the numerically optimized exercise rates (with $k=2$) at $t=0.5$ for $K\in\{100,110\}$.

\begin{figure}[h!]
	\centering
	% This file was created by matplotlib2tikz v0.6.17.
\begin{tikzpicture}

\definecolor{color0}{rgb}{0.12156862745098,0.466666666666667,0.705882352941177}

\begin{axis}[
xmin=82, xmax=138,
ymin=-1.33388983446822, ymax=37.968280468308,
tick align=outside,
tick pos=left,
xlabel={$K$},
x grid style={white!69.01960784313725!black},
y grid style={white!69.01960784313725!black},
xmajorgrids,
xminorgrids,
x grid style={white!50.19607843137255!black},
ymajorgrids,
yminorgrids,
y grid style={white!50.19607843137255!black},
legend style={at={(0.05,0.95)}, anchor=north west, draw=white!80.0!black},
]
\addplot [semithick, color0]%k=2
table {%
80 0.634390633839794
85 1.12311569673604
90 1.86753394828981
95 2.98259701826632
100 4.74272637751241
105 7.06263764316672
110 10.0000000000064
115 15
120 20
125 25
130 30
135 35
140 40
};
\addplot [semithick, color0,style=dashed]%k=1
table {%
70.0 0.19466871013905931
75.0 0.36310912392401473
80.0 0.637173093418396
85.0 1.1040317568641318
89.99999999999999 1.950881936865799
95.0 3.083442925584693
100.0 4.789961309120194
105.0 7.148717419969408
110.00000000000001 10.000000000000007
114.99999999999999 14.999999999999988
120.0 20.0
125.0 25.0
130.0 29.99999999999999
135.0 35.0
140.0 40.0
145.0 45.000000000000014
150.0 50.0
};
\addplot [semithick, color0,style=dotted]%k=0
table {%
70.0 0.1976850463949917
75.0 0.3512733722163651
80.0 0.6223551699232484
85.0 1.0862729512615854
89.99999999999999 1.7480392138902041
95.0 2.7564800337491127
100.0 4.148323215612329
105.0 6.111888036035616
110.00000000000001 10.000000000000007
114.99999999999999 14.999999998996719
120.0 19.99999999892647
125.0 24.99999999870343
130.0 29.999999997929088
135.0 34.99999999888915
140.0 39.99999999871902
145.0 44.99999999873802
150.0 49.99999999776612
};

\addplot [semithick, gray]%European
table {%
80.0 0.6191147233799482
85.00000000000001 1.0374364074187408
90.0 1.7914073908848391
95.0 2.7558865033785143
100.0 4.109627475485303
105.0 6.130031027802647
110.00000000000001 8.524015122083558
114.99999999999999 11.469333750251556
120.0 15.130662787926838
125.0 19.300282091126387
129.99999999999997 23.887837461308013
135.0 28.486956089251137
140.0 33.31720874585505
};

\addlegendentry{ERO ($k=2$)}
\addlegendentry{ERO ($k=1$)}
\addlegendentry{ERO ($k=0$)}
\addlegendentry{European}
\end{axis}

\end{tikzpicture}
	\caption{Dependence of basket put option price on the strike in the $10$-dimensional Heston model.}
	\label{fig:hestonnd}
\end{figure}
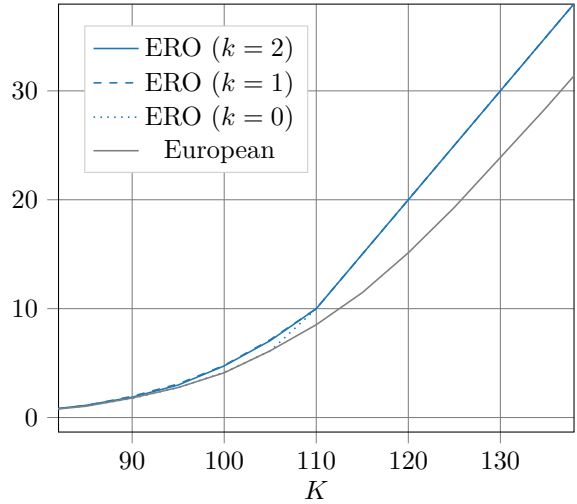

Finally, we consider a 10-dimensional portfolio where each underlying $(X_t^{i})_{t\in\T}, 1\leq i\leq 10$ follows \Cref{hestonunderlying} with the same volatility process $(\variance_t)_{t\in\T}$ (and the same parameter values as in the one-dimensional case) but different Wiener processes $(W^{X^i})_{t\in\T}$, $1\leq i\leq 10$ such that the $11$-dimensional Wiener process $(W^{X^1}_t,\dots,W^{X^{10}}_t,W^{\variance}_t)$ has the covariance matrix
\setcounter{MaxMatrixCols}{20}
\begin{equation*}
\Sigma=\begin{pmatrix}
1.   &  0.2  &  0.2  &  0.35 &  0.2  &  0.25 &  0.2  &  0.2  & 0.3  &  0.2 & -0.5 \\ 
0.2  &  1.   &  0.2  &  0.2  &  0.2  &  0.125&  0.45 &  0.2  & 0.2  &  0.45 & -0.5 \\
0.2  &  0.2  &  1.   &  0.2  &  0.2  &  0.2  &  0.2  &  0.2  & 0.45 &  0.2  & -0.5 \\
0.35 &  0.2  &  0.2  &  1.   &  0.2  &  0.2  &  0.2  &  0.2  & 0.425&  0.2  & -0.5 \\
0.2  &  0.2  &  0.2  &  0.2  &  1.   &  0.1  &  0.2  &  0.2  & 0.5  &  0.2  & -0.5 \\
0.25 &  0.125&  0.2  &  0.2  &  0.1  &  1.   &  0.2  &  0.2  & 0.35 &  0.2  & -0.5 \\
0.2  &  0.45 &  0.2  &  0.2  &  0.2  &  0.2  &  1.   &  0.2  & 0.2  &  0.2  & -0.5 \\
0.2  &  0.2  &  0.2  &  0.2  &  0.2  &  0.2  &  0.2  &  1.   & 0.2  & -0.1  & -0.5 \\
0.3  &  0.2  &  0.45 &  0.425&  0.5  &  0.35 &  0.2  &  0.2  & 1.   &  0.2  & -0.5 \\
0.2  &  0.45 &  0.2  &  0.2  &  0.2  &  0.2  &  0.2  & -0.1  & 0.2  &  1.   & -0.5 \\
-0.5 & -0.5 & -0.5 & -0.5 & -0.5 & -0.5 & -0.5 & -0.5 & -0.5 & -0.5 & 1 \\
\end{pmatrix}
\end{equation*}
\Cref{fig:hestonnd} shows estimates of the values of American basket put options (with coefficients $c\equiv 1/10$) that were obtained by exercise rate optimization for the corresponding $11$-dimensional process $S_t:=(X^{1}_t,\dots,X^{10}_t,\variance_t)$ using the same discretization parameters as before. 

\subsection{Rough volatility}%l=20 for value, l=30 for 
\label{sec:practice3}
To illustrate the wide applicability of our method, we conclude this section with the non-Markovian \emph{rough Bergomi} model, 
which was previously applied to explain implied volatility smiles and other phenomena in the pricing of European options \cite{bayer2016pricing}. In non-Markovian models, \Cref{fund} does not hold because optimal exercise strategies may be based on the entire history of the path $(S_t)_{t\in\T}$, which we again assume to include the underlying asset $(X_t)_{t\in\T}$ as well as the volatility $(\variance_t)_{t\in\T}$. Therefore, we consider the infinite-dimensional Markovian extension
$$
\tilde{S}_t:=(S_u)_{u\in [0,t]},\quad t\in\T,
$$
for which \Cref{fund} formally holds with subsets of $\T\times \R_{+}^{d}$ replaced by subsets of $\T \times \Gamma$, where $\Gamma:=\bigcup_{t\in\T}\{s:[0,t]\to\R_{+}^{d}\}$.  

For numerical purposes, we subsample realizations of $S_t$ (with the convention that $S_t:=S_{0}$ for $t<0$) and define 
$$
\tilde{\bm{S}}_t:=(S_{t},S_{t-\Delta_1},\dots,S_{t-\Delta_J})\in \R^{d_{\text{eff}}}:=\R^{2\times (1+J)},\;\quad t\in \T
$$ 
for some $J<\infty$ and $0<\Delta_1<\dots<\Delta_J$. We apply the algorithm described in \Cref{sec:algo} to the resulting problem of finding exercise rates on the extended space $\T\times \R^{d_{\text{eff}}}$.

Following \cite[Section 4]{bayer2016pricing}, we generate samples from the risk-neutral measure induced by
\begin{align}
	\label{rbunderlying}
	\d X_t &= rX_t\d t+ X_t \sqrt{\variance_{t}} \d W^{X}_t, \quad X_0 = x_0,\\
	\label{rbvolatility}
	\variance_{t}&:=\variance_0\mathcal{E}\left(\eta\sqrt{2H}\int_{0}^{t}\frac{1}{(t-u)^{1/2-H}}\d W^{\variance}_{u}\right),
\end{align}
where $\mathcal{E}$ is the stochastic exponential in the Wick sense, $H = 0.07$, $r=0.05$, $\eta = 1.9$, and  $W^{X}$, $W^{\variance}$ are Wiener processes with correlation $\rho=-0.9$.
Since the asset price process $X_t$ is a continuous local martingale, standard no arbitrage theory applies even though $\variance_t$ is not a semi-martingale.

\Cref{table:bergomi} shows the American option prices for $x_0=100$, $\variance_0=0.09$, $T=1$, and different strikes, which we computed using the discretization parameters $M=\SI{100000}{}$, $N=128$, $k=2$, and  $\Delta_j:=j/8$, $1\leq j\leq J$, $J\in\{0,1,3,7\}$. For comparison, we include the European prices computed by simple Monte Carlo simulation. 
The difference between our estimates for $J=0$ and $J=7$ is not consistently larger than the Monte Carlo sampling error, indicating that the exploitation of non-Markovian features does not yield significantly improved exercise strategies. This is not to say, however, that American option prices in non-Markovian and Markovian models are similar. The non-Markovianity of the samples of $(S_t)_{t\in\T}$ plays an important role in the evaluation of any given strategy, even when the strategy only depends on the spot values.

\begin{table}[h!]
	\centering
	\begin{tabular}{rc|cccccccc}
		& & \multicolumn{8}{c}{$K$}\\
	& & $70$ & $80$ & $90$ & $100$ & $110$ & $120$ & $130$ & $140$\\ 
	\hline
\multicolumn{2}{c|}{Euro.} & 1.83&3.13&5.06&7.98&12.21&17.99&25.35&33.88\\
& 0 & 1.88&3.23&5.32&8.51&13.24&20&30&40\\
  & 1 &1.88& 3.23&5.31&8.50&13.22&20&30&40\\
 \raisebox{0.5em}[0pt]{J} &%
 3   &1.88&3.21& 5.31&8.50&13.22&20&30&40\\
 &7 &1.88&3.22& 5.30&8.50&13.23& 20 & 30 &40
	\end{tabular}
	\caption{Prices of put options in the rough Bergomi model.}
	\label{table:bergomi}
\end{table}

\begin{figure}[h!]
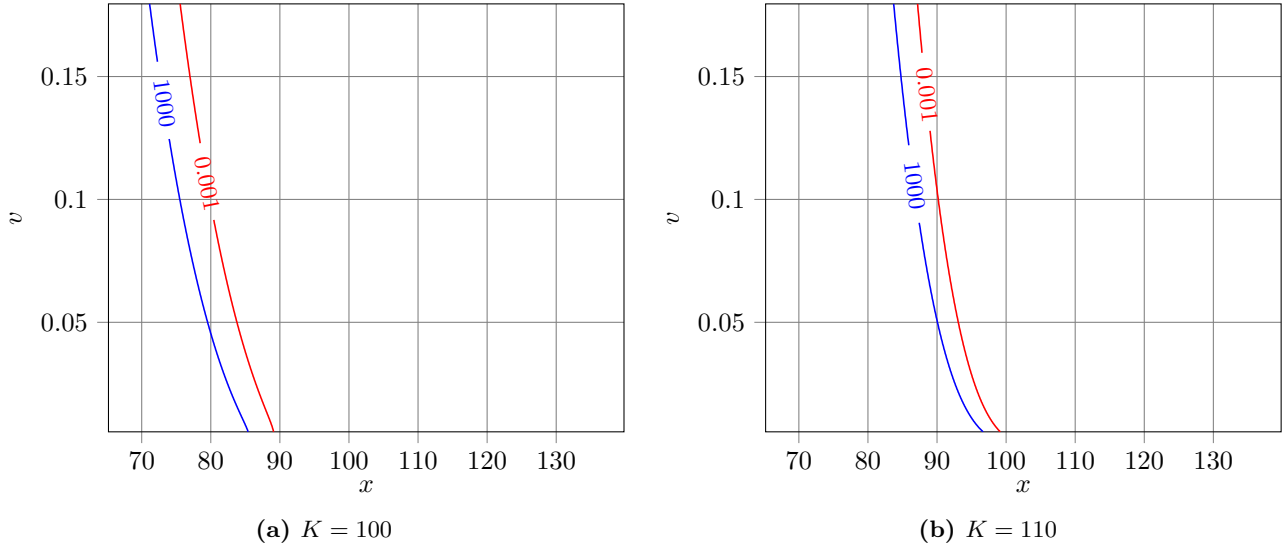

	\begin{subfigure}{0.49\textwidth}
		\input{figures/contourbergomi100}
		\caption{$K=100$}
	\end{subfigure}
	\begin{subfigure}{0.5\textwidth}
		\input{figures/contourbergomi110}
		\caption{$K=110$}
	\end{subfigure}
	\caption{Level sets of optimal exercise rates at $t=0.5$ for put options in the rough Bergomi model ($k=2$, $J=0$).}
	\label{fig:exratebergomi}
\end{figure}

The numerically optimized exercise rates at $t=0.5$ for $J=0$ and  $K\in\{100,110\}$ are shown in \Cref{fig:exratebergomi}.

\section{Conclusion}

We have introduced a method of pricing American options by optimization of randomized exercise strategies, in which deterministic exercise regions are replaced by probabilistic exercise rates. 

Since the objective function of the corresponding relaxed optimization problem is smooth, optimal exercise rates can be found using simple deterministic optimization routines. 
Our numerical experiments show that exercise rates based on quadratic polynomials are sufficient to obtain remarkably accurate price estimates and that the resulting non-concave objective functions can be globally maximized using only a few iterations. Since the market model only appears in the simulation of sample paths, our method is quite flexible and easy to implement. We demonstrated its practical applicability in uni- and multivariate Black--Scholes, Heston and rough Bergomi models. 

In even higher-dimensional situations than those considered in this work, already the space of quadratic polynomials may be prohibitively large. In that case, the polynomial subspace $\mathcal{P}$ could be designed in an anisotropic way to exploit, for example, the fact that the exercise decision of basket put options with coefficients $c$ is most sensitive to the coordinate $\tilde{s}_1:=c\cdot s$.
For situations where large polynomial subspaces are unavoidable, a rigorous analysis of the number of samples that are required to determine a given number of degrees of freedom without significant overfitting would be of interest; similar but not directly transferable results were established in \cite{belomestny2011rates,zanger2018convergence}. 

To accelerate numerical implementations, multilevel Monte Carlo methods \cite{Giles2015} could be used for evaluations of the expected payoff and its gradient. 

It is an open question whether efficiently computable upper bounds on the option price \cite{belomestny2013solving} can be constructed using exercise rates as well.

{\small 
\textbf{Acknowledgments}
This work was supported by the KAUST Office of Sponsored Research (OSR, award URF/1/2584-01-01), the German Research Foundation (DFG, grant BA5484/1) and the Alexander von Humboldt Foundation. R. Tempone and S. Wolfers are members of the KAUST SRI Center for Uncertainty Quantification in Computational Science and Engineering.
}

\begin{thebibliography}{10}

\bibitem{achdou2005computational}
Yves Achdou and Olivier Pironneau.
\newblock {\em Computational methods for option pricing}.
\newblock SIAM, 2005.

\bibitem{andersen1999simple}
Leif Andersen.
\newblock A simple approach to the pricing of {B}ermudan swaptions in the
  multi-factor {LIBOR} market model.
\newblock {\em Journal of Computational Finance}, 3:5--32, 1999.

\bibitem{andersen2004primal}
Leif Andersen and Mark Broadie.
\newblock Primal-dual simulation algorithm for pricing multidimensional
  {A}merican options.
\newblock {\em Management Science}, 50(9):1222--1234, 2004.

\bibitem{BagbyBosLevenberg2002}
Thomas Bagby, Len Bos, and Norman Levenberg.
\newblock Multivariate simultaneous approximation.
\newblock {\em Constructive Approximation}, 18(4):569, December 2002.

\bibitem{barone1987efficient}
Giovanni Barone-Adesi and Robert~E. Whaley.
\newblock Efficient analytic approximation of {A}merican option values.
\newblock {\em The Journal of Finance}, 42(2):301--320, 1987.

\bibitem{bayer2016pricing}
Christian Bayer, Peter Friz, and Jim Gatheral.
\newblock Pricing under rough volatility.
\newblock {\em Quantitative Finance}, 16(6):887--904, 2016.

\bibitem{bellman2015adaptive}
Richard {Bellman}.
\newblock {\em {Adaptive control processes: A guided tour. (A RAND Corporation
  Research Study).}}
\newblock Princeton University Press, 1961.

\bibitem{belomestny2011on}
Denis Belomestny.
\newblock On the rates of convergence of simulation-based optimization
  algorithms for optimal stopping problems.
\newblock {\em Ann. Appl. Probab.}, 21(1):215--239, 02 2011.

\bibitem{belomestny2011rates}
Denis Belomestny et~al.
\newblock On the rates of convergence of simulation-based optimization
  algorithms for optimal stopping problems.
\newblock {\em The Annals of Applied Probability}, 21(1):215--239, 2011.

\bibitem{belomestny2013solving}
Denis Belomestny et~al.
\newblock Solving optimal stopping problems via empirical dual optimization.
\newblock {\em The Annals of Applied Probability}, 23(5):1988--2019, 2013.

\bibitem{belomestny2018advanced}
Denis Belomestny and John Schoenmakers.
\newblock {\em Advanced Simulation-Based Methods for Optimal Stopping and
  Control: With Applications in Finance}.
\newblock Springer, 2018.

\bibitem{broadie1997valuation}
Mark Broadie and J{\'e}r{\^o}me Detemple.
\newblock The valuation of {A}merican options on multiple assets.
\newblock {\em Mathematical Finance}, 7(3):241--286, 1997.

\bibitem{broadie1997pricing}
Mark Broadie and Paul Glasserman.
\newblock Pricing {A}merican-style securities using simulation.
\newblock {\em Journal of Economic Dynamics and Control}, 21(8):1323 -- 1352,
  1997.

\bibitem{byrd1995limited}
Richard~H Byrd, Peihuang Lu, Jorge Nocedal, and Ciyou Zhu.
\newblock A limited memory algorithm for bound constrained optimization.
\newblock {\em SIAM Journal on Scientific Computing}, 16(5):1190--1208, 1995.

\bibitem{gobet2006sensitivities}
Cristina Costantini, Emmanuel Gobet, and Nicole El~Karoui.
\newblock Boundary sensitivities for diffusion processes in time dependent
  domains.
\newblock {\em Appl. Math. Optim.}, 54(2):159--187, 2006.

\bibitem{cox1979option}
John~C. Cox, Stephen~A. Ross, and Mark Rubinstein.
\newblock Option pricing: A simplified approach.
\newblock {\em Journal of financial Economics}, 7(3):229--263, 1979.

\bibitem{garcia2003convergence}
Diego Garc{\i}a.
\newblock Convergence and biases of {M}onte {C}arlo estimates of {A}merican
  option prices using a parametric exercise rule.
\newblock {\em Journal of Economic Dynamics and Control}, 27(10):1855--1879,
  2003.

\bibitem{gemmrich2012master}
Simon Gemmrich.
\newblock Multilevel {M}onte {C}arlo methods for {A}merican options.
\newblock Master's thesis, University of Oxford, 2012.

\bibitem{Giles2015}
Michael~B. Giles.
\newblock Multilevel {M}onte {C}arlo methods.
\newblock {\em Acta Numerica}, 24:259--328, 2015.

\bibitem{grant1997path}
Dwight Grant, Gautam Vora, and David Weeks.
\newblock Path-dependent options: Extending the {M}onte {C}arlo simulation
  approach.
\newblock {\em Management Science}, 43(11):1589--1602, 1997.

\bibitem{gyongy2008randomized}
Istv{\'a}n Gy{\"o}ngy and David {\v{S}}i{\v{s}}ka.
\newblock On randomized stopping.
\newblock {\em Bernoulli}, 14(2):352--361, 2008.

\bibitem{heston1993closed}
Steven~L Heston.
\newblock A closed-form solution for options with stochastic volatility with
  applications to bond and currency options.
\newblock {\em The review of financial studies}, 6(2):327--343, 1993.

\bibitem{ibanez2004monte}
Alfredo Ibanez and Fernando Zapatero.
\newblock {M}onte {C}arlo valuation of {A}merican options through computation
  of the optimal exercise frontier.
\newblock {\em Journal of Financial and Quantitative Analysis}, 39(2):253--275,
  2004.

\bibitem{karatzas1998methods}
Ioannis Karatzas and Steven~E. Shreve.
\newblock {\em Methods of mathematical finance}.
\newblock Springer, 1998.

\bibitem{krylov2008controlled}
Nikolaj~Vladimirovi{\v{c}} Krylov.
\newblock {\em Controlled diffusion processes}, volume~14.
\newblock Springer Science \& Business Media, 2008.

\bibitem{kuske1998optimal}
Rachel~A. Kuske and Joseph~B. Keller.
\newblock Optimal exercise boundary for an {A}merican put option.
\newblock {\em Applied Mathematical Finance}, 5(2):107--116, 1998.

\bibitem{longstaff2001valuing}
Francis~A. Longstaff and Eduardo~S. Schwartz.
\newblock Valuing {A}merican options by simulation: a simple least-squares
  approach.
\newblock {\em Review of Financial Studies}, 14(1):113--147, 2001.

\bibitem{ludkovski2018kriging}
Michael Ludkovski.
\newblock Kriging metamodels and experimental design for {B}ermudan option
  pricing.
\newblock {\em Journal of Computational Finance}, 22(1):37--77, 2018.

\bibitem{reisinger2007efficient}
Christoph Reisinger and Gabriel Wittum.
\newblock Efficient hierarchical approximation of high-dimensional option
  pricing problems.
\newblock {\em SIAM Journal on Scientific Computing}, 29(1):440--458, 2007.

\bibitem{rogers2002montecarlo}
Leonard C.~G. Rogers.
\newblock {M}onte {C}arlo valuation of {A}merican options.
\newblock {\em Mathematical Finance}, 12(3):271--286, 2002.

\bibitem{shiryaev2007optimal}
Albert~N. Shiryaev.
\newblock {\em Optimal stopping rules}, volume~8.
\newblock Springer Science \& Business Media, 2007.

\bibitem{zanger2018convergence}
Daniel~Z. Zanger.
\newblock Convergence of a least-squares {M}onte {C}arlo algorithm for
  {A}merican option pricing with dependent sample data.
\newblock {\em Mathematical Finance}, 28(1):447--479, 2018.

\end{thebibliography}
\end{document}